\documentclass[12pt]{article}

\usepackage{amsmath,amssymb,amsfonts,amscd,mathrsfs,mathtools,bbold, braket}
\usepackage{xcolor}
\definecolor{darkblue}{rgb}{0.1,0.1,.7}
\usepackage[colorlinks, linkcolor=darkblue, citecolor=darkblue, urlcolor=darkblue, linktocpage]{hyperref} 
\usepackage{geometry}
\usepackage{subcaption}
\usepackage{tablefootnote,colortbl}

\usepackage{ physics }

\usepackage{braket}
\usepackage{cancel}
\usepackage{arcs}

\usepackage{arydshln}
\usepackage{booktabs,multirow}
\usepackage{amssymb}
\usepackage{fancyvrb}

\usepackage{tikz-feynman}
 \tikzfeynmanset{compat=1.0.0}

 \usepackage{mcite}
 
 \usepackage[square, comma, compress,numbers]{natbib}
\usepackage{subcaption}
\usepackage[utf8]{inputenc} 

\definecolor{myorange}{RGB}{199,146,32}

\definecolor{Gray1}{gray}{0.97}
\definecolor{Gray2}{gray}{0.9}
\definecolor{LightCyan}{rgb}{0.88,1,1}
\definecolor{blu}{rgb}{0,0,1}

\usepackage{ragged2e}
\usepackage{array}
\newcolumntype{L}[1]{>{\raggedright\let\newline\\\arraybackslash\hspace{0pt}}m{#1}}
\newcolumntype{C}[1]{>{\centering\let\newline\\\arraybackslash\hspace{0pt}}m{#1}}
\newcolumntype{R}[1]{>{\raggedleft\let\newline\\\arraybackslash\hspace{0pt}}m{#1}}

\usepackage{dsfont} 
\usepackage{tcolorbox}
\usepackage{booktabs}

\usepackage{titlesec}
\titleformat*{\section}{\large\bfseries}
\titleformat*{\subsection}{\normalsize\bfseries}
\titleformat*{\subsubsection}{\normalsize\it}
\titleformat*{\paragraph}{\normalsize\bfseries}
\titleformat*{\subparagraph}{\normalsize\bfseries}

\newcommand{\reef}[1]{(\ref{#1})}

\newcommand{\beq}{\begin{equation}} 
\newcommand{\eeq}{\end{equation}}

\newcommand{\diffop}[2]{\ifthenelse{\equal{#2}{1}}{\frac{\mrm{d}}{\mrm{d} #1}}{\frac{\mrm{d}^#2}{\mrm{d} #1^#2}}}

\newcommand{\bk}{{\mathbf k}}

\newcommand{\s}[1]{\langle #1 \rangle}

\newcommand{\mrm}[1]{{\mathrm #1}}

\newcommand{\re}{\text{Re}\, }

\def\tr{\mathrm{tr}}
\usepackage[normalem]{ulem}

\newcommand{\be}{\begin{equation}}
\newcommand{\ee}{\end{equation}}
\def\bea#1\eea{\begin{align}#1\end{align}}

\newcommand{\eq}[1]{Eq.~(\ref{#1})}

\definecolor{Ecolor}{RGB}{106,157,235}

\usepackage{accents}
\newlength{\dhatheight}

\usepackage{tikz}
\usetikzlibrary{arrows,shapes}
\usetikzlibrary{trees}
\usetikzlibrary{matrix,arrows} 				
\usetikzlibrary{calc} 
\usetikzlibrary{positioning}				
\usetikzlibrary{calc,through}				
\usetikzlibrary{decorations.pathreplacing}  
\usepackage[tikz]{bclogo} 					
\usepackage{pgffor}							

\usetikzlibrary{decorations.pathmorphing}	
\usetikzlibrary{decorations.markings}
\usetikzlibrary{intersections}				
\tikzset{
    vector/.style={decorate, decoration={snake}, draw},
    graviton/.style={decorate, decoration={snake,amplitude=1.5pt}, draw},
    fermion/.style={postaction={decorate},
        decoration={markings,mark=at position .55 with {\arrow{>}}}},
    fermionbar/.style={draw, postaction={decorate},
        decoration={markings,mark=at position .55 with {\arrow{<}}}},
    fermionnoarrow/.style={},
    gluon/.style={decorate,
        decoration={coil,amplitude=4pt, segment length=5pt}},
    scalar/.style={dashed, postaction={decorate},
        decoration={markings,mark=at position .55 with {\arrow{>}}}},
    scalarbar/.style={dashed, postaction={decorate},
        decoration={markings,mark=at position .55 with {\arrow{<}}}},
    scalarnoarrow/.style={dashed,draw},
	provector/.style={decorate, decoration={snake,amplitude=2.5pt}, draw},
	antivector/.style={decorate, decoration={snake,amplitude=-2.5pt}, draw},
	    electron/.style={draw=black, postaction={decorate},
        decoration={markings,mark=at position .55 with {\arrow[draw=black]{>}}}},
	bigvector/.style={decorate, decoration={snake,amplitude=4pt}, draw},
	vectorscalar/.style={loosely dotted,draw=black, postaction={decorate}},
}

\numberwithin{equation}{section}
\usepackage{tocloft}
\renewcommand{\order}[1]{\ensuremath{O\left(#1\right)}}

\begin{document}

\vspace*{-.6in} \thispagestyle{empty}
\begin{flushright}
\end{flushright}
\vspace{1cm} {\Large
\begin{center}
\textbf{
Thermodynamics from the S-matrix reloaded,\\[0.2cm] with applications to QCD and the  confining   Flux Tube
 }
\end{center}}
\vspace{1cm}
\begin{center}

{\bf  Pietro Baratella$^{a}$,  Joan Elias~Mir\'o$^{b}$, Emanuele  Gendy$^{c}$ } \\[1cm] 
 {$^a$  Institut Jo\v{z}ef Stefan, Jamova cesta 39, 1000, Ljubljana, Slovenija  \\ 
  $^b$ The Abdus Salam ICTP,    Strada Costiera 11, 34151, Trieste, Italy  \\ 
   $^c$  Technische Universit\"{a}t M\"{u}nchen, Physik-Department, 85748 Garching, Germany  \\ 
 }
\vspace{1cm}

\abstract{ 

\bigskip
\noindent

Over the past decade and more, $S$-matrix-based calculational methods have experienced a resurgence, proving to be both an elegant and powerful tool for extracting physical quantities without the need for an underlying Lagrangian formulation. In this work, we critically review and further develop the formalism introduced by Dashen, Ma, and Bernstein, which connects the thermodynamics of relativistic systems to the information contained in their scattering amplitudes. As a demonstration, we revisit the computation of the QCD equation of state to leading order in the strong coupling, showcasing the advantages of this method over traditional Thermal Field Theory techniques. Additionally, we apply these tools to the effective theory of a long confining  Flux Tube in $D$ dimensions, analyzing thermal effects up to and including NNLO contributions. Our results are compared with those obtained using  the Thermodynamic Bethe Ansatz method. We also discuss how these techniques enable the inclusion of non-universal  effects in the study of Flux Tubes, while relying solely on the $S$-matrix as input.

}

\vspace{3cm}
\end{center}

 \vfill
 {
  \flushright
 \today 
}

\newpage 

\setcounter{tocdepth}{2}

{
\tableofcontents
}
 
 

\section{Introduction}

More than half a century ago R.~Dashen, S.-K.~Ma and H.~J.~Bernstein~(DMB)~\cite{Dashen:1969ep,dashen1970singular} uncovered a beautiful expression for the thermal partition function $Z(\beta)\equiv \Tr e^{-\beta H}$ of a system in terms of its  $S$-matrix elements, reading
\be\label{masterformula}
Z(\beta) =Z_{0}(\beta)+\frac{\beta}{2\pi i}\int_0^\infty  \dd{E} \, e^{-\beta E} \,  \Tr \ln S(E) \, , 
\ee
where $\beta=(k_BT)^{-1}$, $Z_0$ is the partition function in the non-interacting limit and $S(E)$ is an operator-valued distribution in the real energy variable $E$ with the property that
\be\label{S_onshell}
\langle \beta | S(E=E_\alpha) | \alpha \rangle = S_{\beta\alpha}\,,
\ee
with $S_{\beta\alpha}$ the ordinary $S$-matrix element for the transition from the incoming state $\alpha$ to the outgoing state $\beta$. Turning on a chemical potential in \reef{masterformula} is straightforward and  amounts to appropriately weigh the states in the trace by the corresponding charge. 

The free energy of the system is given by $F(\beta)\equiv-\beta^{-1} \ln Z(\beta)$, and is obtained by keeping only the extensive
contribution to the partition function 
\be
F(\beta) =F_{0}(\beta)- \frac{1}{2\pi i}\int_0^\infty  \dd{E} \, e^{-\beta E} \,  \Tr_c \ln S(E)\, ,
\label{master}
\ee
where  
the subscript $c$ indicates that, once the formula is expanded in a diagrammatic fashion, we should only keep terms corresponding to diagrams that remain connected after the trace is taken. We will call these terms \emph{trace-connected}, or simply \emph{connected} when there is no ambiguity. 
The operator $S(E)$ is unitary and therefore $\Tr \ln S(E)$ is  purely imaginary.

Let us sketch  the derivation of Eq. \reef{master}, which will be at the basis of all our calculations. The key qualitative observation is that both the partition functions of the free and interacting theory and the operator $S(E)$ are expressible in terms of resolvents, $G(z)=(z-H)^{-1}$, where $z$ is a complex energy, with a similar expression for the free-theory resolvent $G_0(z)$. Specifically we have $Z(\beta)=-\frac{1}{\pi} \lim_{\epsilon \to 0^+}\int_{0}^{\infty} {\dd E} e^{-\beta E} \,{\rm Im} \,\Tr G(E+i\epsilon)$, and similarly for $Z_0$, while $S(E)\equiv \lim_{\epsilon\to 0^+}\Omega^{-1}(E-i\epsilon)\Omega(E+i\epsilon)$, with $\Omega(z)= G(z)G_0^{-1}(z)$. The identity involving the partition function is obtained with standard contour methods starting from $\Tr e^{-\beta H}$, while the second identity should be taken for the moment just as the definition of $S(E)$.
From this definitions it is  an algebraic exercise  to obtain the key identity 
 $
\partial_E \Tr \ln S(E)=-2i\Im \Tr (G(E+i\epsilon)-G_0(E+i\epsilon)),
$
 whose derivation can be found in \cite{Dashen:1969ep}.  
From the last identity it is straightforward to  derive  \reef{masterformula}, and \reef{master} as the connected version of it.

From the physics point of view, it only remains to establish the connection, anticipated in \reef{S_onshell}, among $S(E)$ and the usual $S$-matrix we deal with in everyday life.  From Scattering Theory   it follows that
$S(E)=\mathbb{1}-2\pi i \delta(E-H_0)T(E+i\epsilon)$, where $T(z)=(1-VG_0(z))^{-1}V$ and $V=H-H_0$ is the interaction Hamiltonian. The transmission part $T(E+i\epsilon)$ of $S(E)$, once expanded in powers of $V$ and bracketed with asymptotic states $\alpha\to\beta$,  can be expressed as
\be\label{T_OFPT}
T_{\beta\alpha}(E)=V_{\beta\alpha}+\int \dd \gamma \,\frac{V_{\beta\gamma}V_{\gamma\alpha}}{E-E_\gamma+i\epsilon}+\int \dd \gamma\dd\gamma' \,\frac{V_{\beta\gamma}V_{\gamma\gamma'}V_{\gamma'\alpha}}{(E-E_\gamma+i\epsilon)(E-E_{\gamma'}+i\epsilon)}+\ldots
\ee
where $\int \dd \gamma$ denotes a sum over intermediate states that resolve the product of operators. 
When $T_{\beta\alpha}(E)$ is evaluated at $E{=}E_\alpha$ the above equation becomes the Lippmann-Schwinger equation of Old-Fashioned Perturbation Theory (OFPT);
 and when combined with the $\delta(E-E_\beta)$ support and the non-interacting part $\mathbb{1}$, equation \reef{S_onshell} follows. 

Structurally, \reef{master} relates a zero-temperature property of the given system, specifically its scattering matrix, to the observables of the system when heated at temperature $\beta^{-1}$. Recent progress in on-shell Scattering  amplitude methods have lead  to a deeper understanding of the structure of perturbative QCD scattering amplitudes and to an improvement in our ability to compute them at higher loop orders and external particle multiplicities. This wealth of knowledge could be in principle used, following the instructions of \reef{master}, to compute to high orders the (low $\beta$) QCD equation of state.

Another interesting regime of QCD is that of confining Flux Tubes of length $R=\beta$, which can be studied as an effective field theory (EFT) of long strings, a 1+1 dimensional QFT. The ground state energy of a Flux Tube as a function of $R$ can be computed with lattice Monte Carlo  methods starting from the QCD Lagrangian. It can also be extracted from the free energy $F(R)$ of the long string EFT (in fact the two quantities are the same) once an effective Lagrangian is specified, so that lattice results can in principle be used to constrain the EFT coefficients according to how they enter in $F(R)$. For the task of extracting $F(R)$, Eq.~\reef{master} stands as a valuable alternative to Thermodynamic Bethe Ansatz (TBA) techniques that have been applied over the past decade, especially when studying non-integrable deformations (of the integrable realization) of Flux Tubes. Common to the two methods is that they use knowledge of scattering amplitudes to deduce thermodynamic properties. However the DMB method is more general, in that it does not assume integrability of the 1+1 dimensional theory.

More in general, we believe that recent development in the understanding of scattering amplitudes has made this the right time to refresh an elegant formalism that was, in our opinion, not developed to a satisfactory level over the past, especially in terms of concrete examples of its application.

The paper is structured as follows. In Sec.~\ref{sec:QCD} we discuss the diagrammatics of \reef{master} and consider the leading deviations from the free-theory partition function, studying the QCD equation of state at $O(\alpha_s)$ as an example. In Sec.~\ref{sec:fluxtuberecap} we study the integrable realization of a long Flux Tube, extracting its free energy up to NLO and comparing it with the exact result previously obtained with TBA methods. Sec.~\ref{sec:higherorders} is devoted to the study of NNLO effects, after which we comment on the similarities and differences with respect to the TBA method, and then show how the formula can be used to directly include contributions that break
the integrability of the system. Sec.~\ref{sec:conclusion} presents our conclusions.

\section{Setting  the stage: the QCD equation of state at $O(\alpha_s)$}\label{sec:QCD}

To delve deeper into the application of   \reef{master} we are going to explain the calculation 
of the QCD thermal partition function up to $O(g_s^2)$. 
This will highlight the power of the on-shell method  based on \reef{master} compared to the  standard textbook technique based on thermal  Feynman diagrams.

Eq. \reef{master} suggests a perturbative expansion of the free energy in terms of $S$-matrix elements with increasing number of scattering particles. Dividing the operator $S(E)$ into a transmission piece $\mathbb{1}$ and an interacting $T$-matrix,
the logarithm in \reef{master} implies that we should also include traces over successive rescatterings, as dictated by the expansion
\be
\ln S(E)=-\sum_{N=1}^\infty\frac{1}{N}\left[2\pi i \delta(E-H_0)T(E+i\epsilon)\right]^{N} \, .  \label{pertT}
\ee
To compute traces, we bracket over a complete set of Fock-space states, as
\bea\label{traceO}
\Tr  O  =\sum_{n=0}^\infty \frac{1}{n!}\bigg(\prod_{i=1}^n\int\frac{\dd[d]{k_i}}{2E_i(2\pi)^d}\bigg)\bra{\bk_1,\dots,\bk_n} O \ket{\bk_1,\dots,\bk_n}\ ,
\eea
with single particle states being normalized accordingly as $\bra{\bk}\ket{\bk'}=2E_{\bk}(2\pi)^d\delta^{(d)}(\bk-\bk')$, and where $d$ is the number of space dimensions. In order to resolve operator products we will insert identities as sums over Fock-space states. It is sometimes convenient to write \reef{traceO} schematically as $\Tr O=\int \dd \alpha\,O_{\alpha\alpha}$, with $\alpha$ running over a complete set of states.

First order of business is to compute the \emph{free} part of the free energy.
The value of this calculation is to clarify the meaning of \emph{trace-connectedness}. 
The logarithm of the free-theory partition function can indeed be computed by restricting to trace-connected contributions, that is $F_0(\beta)=-\beta^{-1}\Tr_c  \exp(-\beta H_0)$. Given that the free Hamiltonian is diagonal in the Fock basis, we have  to evaluate  a sum over
\be
\bra{ \bk_1,\ldots,\bk_n} \ket{\bk_1,\ldots,\bk_n}_c\,, \label{nnorm}
\ee
weighted by $\prod_{i=1}^n\exp(-\beta E_i)$. 
Eq.~\reef{nnorm}, which is nothing but the inner product of a state with itself,  can be thought of as a forward amplitude in the free-theory. 
This amplitude is not connected in the usual sense because there are no interaction vertices. 
However, when identical particles are present, the free-theory amplitude can be trace-connected. 
For instance, the following picture shows examples of a trace-connected four-particle history (left) and of a trace-\emph{dis}connected four-particle history (right). 
\begin{figure}[h]\centering
\includegraphics[width=10cm]{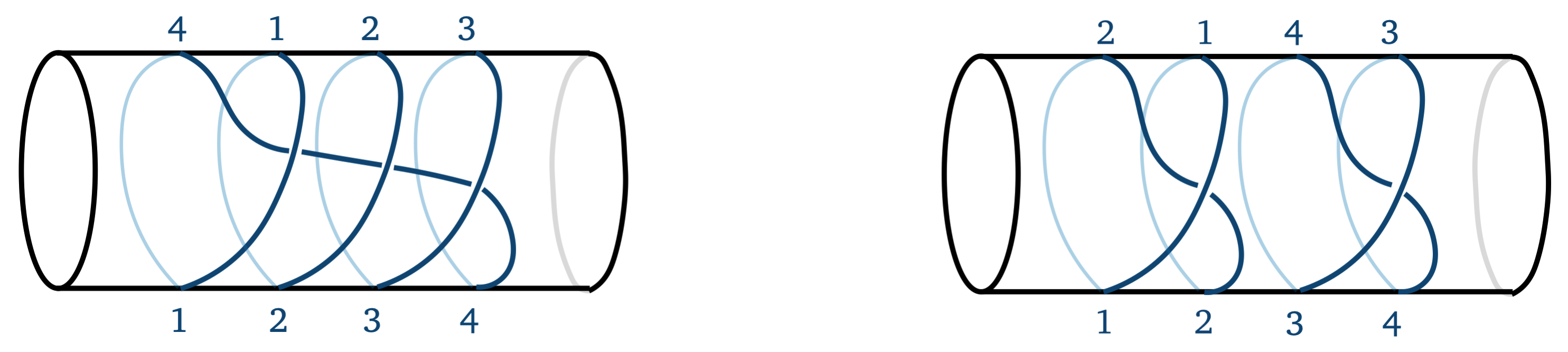}
\end{figure}

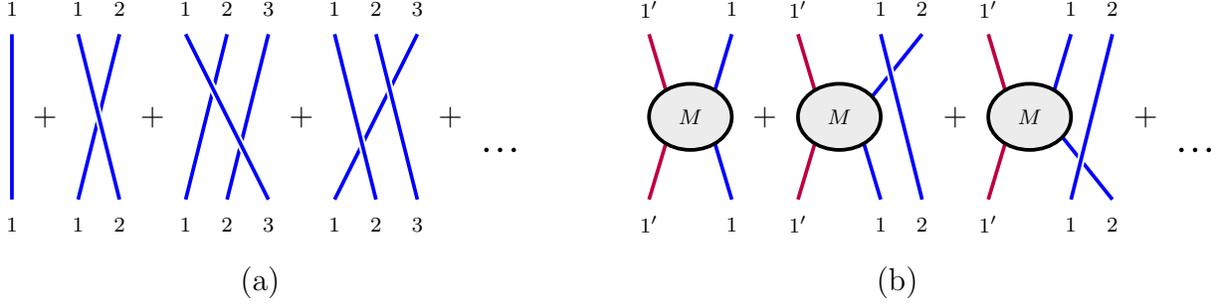
\begin{figure}[t]
	\centering
	\begin{tikzpicture}[line width=1.4 pt, scale=1.1, baseline=(current bounding box.center)]

	\draw[blue] (-10,-1) -- (-10,1) ;
	\node at (-10,-1.3) {\scriptsize 1} ;
	\node at (-10,1.3) {\scriptsize 1} ;
	
	\node at (-9-.6,0) {+} ;
	
	\begin{scope}[shift={(-1.2,0)}]
	\draw[blue] (-8,-1) -- (-7.5,1) ;
	\draw[white, line width= 3.] (-7.5,-1) -- (-8,1) ;
	\draw[blue] (-7.5,-1) -- (-8,1) ;
	\node at (-8,-1.3) {\scriptsize 1} ;
	\node at (-8,1.3) {\scriptsize 1} ;
	\node at (-7.5,-1.3) {\scriptsize 2} ;
	\node at (-7.5,1.3) {\scriptsize 2} ;
	\end{scope} ;
	
	\node at (-6.5-1.8,0) {+} ;
	
	\begin{scope}[shift={(-2.4,0)}]
	\draw[blue] (-5.5,-1) -- (-5,1) ;
	\draw[blue] (-5,-1) -- (-4.5,1) ;
	\draw[white, line width= 3.] (-4.5,-1) -- (-5.5,1) ;
	\draw[blue] (-4.5,-1) -- (-5.5,1) ;
	\node at (-5.5,-1.3) {\scriptsize 1} ;
	\node at (-5.5,1.3) {\scriptsize 1} ;
	\node at (-5,-1.3) {\scriptsize 2} ;
	\node at (-5,1.3) {\scriptsize 2} ;
	\node at (-4.5,-1.3) {\scriptsize 3} ;
	\node at (-4.5,1.3) {\scriptsize 3} ;
	\end{scope} ;
	
	\node at (-3.5-3,0) {+} ;
	
	\begin{scope}[shift={(-3.6,0)}]
	\draw[blue] (-2.5,-1) -- (-1.5,1) ;
	\draw[white, line width= 3.] (-2,-1) -- (-2.5,1) ;
	\draw[blue] (-2,-1) -- (-2.5,1) ;
	\draw[white, line width= 3.] (-1.5,-1) -- (-2,1) ;
	\draw[blue] (-1.5,-1) -- (-2,1) ;
	\node at (-2.5,-1.3) {\scriptsize 1} ;
	\node at (-2.5,1.3) {\scriptsize 1} ;
	\node at (-2,-1.3) {\scriptsize 2} ;
	\node at (-2,1.3) {\scriptsize 2} ;
	\node at (-1.5,-1.3) {\scriptsize 3} ;
	\node at (-1.5,1.3) {\scriptsize 3} ;
	\end{scope} ;
	
	\node at (-.5-4.2,0) {+} ;
	
	\node at (.7-4.8,-.4) {\LARGE{...}} ;	
	
	\node at (-7,-2.) {(a)} ;	
	
	
	\begin{scope}[shift={(-5.8,0)}]
	
	\draw[purple] (3.5,-1) -- (3.8,0) -- (3.5,1) ;
	\draw[blue] (4.5,-1) -- (4.2,0) -- (4.5,1) ;
	\node at (3.5,-1.3) {\scriptsize $1'$} ;
	\node at (3.5,1.3) {\scriptsize $1'$} ;
	\node at (4.5,-1.3) {\scriptsize 1} ;
	\node at (4.5,1.3) {\scriptsize 1} ;
	
	\draw[fill=gray!15] (4,0) ellipse (.5 cm and .4 cm) ;
	\node at (4,0) {\scriptsize $M$} ;
	
	\node at (5.5-.6,0) {+} ; 
	
	\begin{scope}[shift={(-1.2,0)}]
	\draw[purple] (6.5,-1) -- (6.8,0) -- (6.5,1) ;
	\draw[blue] (7.5,-1) -- (7.2,0) -- (8,1) ;
	\draw[white, line width = 3.] (8,-1) -- (7.5,1) ;
	\draw[blue] (8,-1) -- (7.5,1) ;
	\node at (6.5,-1.3) {\scriptsize $1'$} ;
	\node at (6.5,1.3) {\scriptsize $1'$} ;
	\node at (7.5,-1.3) {\scriptsize 1} ;
	\node at (7.5,1.3) {\scriptsize 1} ;
	\node at (8,-1.3) {\scriptsize 2} ;
	\node at (8,1.3) {\scriptsize 2} ;
	
	\draw[fill=gray!15] (7,0) ellipse (.5 cm and .4 cm) ;
	\node at (7,0) {\scriptsize $M$} ;
	\end{scope};
	
	\node at (9-1.8,0) {+} ; 
	
	\begin{scope}[shift={(-2.4,0)}]
	\draw[purple] (10,-1) -- (10.3,0) -- (10,1) ;
	\draw[blue] (11.5,-1) -- (10.7,0) -- (11,1) ;
	\draw[white, line width = 3.] (11,-1) -- (11.5,1) ;
	\draw[blue] (11,-1) -- (11.5,1) ;
	\node at (10,-1.3) {\scriptsize $1'$} ;
	\node at (10,1.3) {\scriptsize $1'$} ;
	\node at (11,-1.3) {\scriptsize 1} ;
	\node at (11,1.3) {\scriptsize 1} ;
	\node at (11.5,-1.3) {\scriptsize 2} ;
	\node at (11.5,1.3) {\scriptsize 2} ;
	
	\draw[fill=gray!15] (10.5,0) ellipse (.5 cm and .4 cm) ;
	\node at (10.5,0) {\scriptsize $M$} ;
	\end{scope} ;
	
	\node at (12.5-3,0) {+} ; 
	
	\node at (13.7-3.6,-.4) {\LARGE{...}} ;	
	
	\node at (6.5,-2.) {(b)} ;	
	
	\end{scope};

	\end{tikzpicture}
	\caption{\emph{{\rm (a)} Trace-connected histories with $n$=1,2,3, contributing to the free-theory part of the free energy. For a given $n$, the $(n-1)!$ distinct histories are in one-to-one correspondence with the cyclic permutations of $\{1,2,\ldots, n \}$. {\rm (b)} Some contributions to \eq{2to2}, having $n$=1 and $m$=1,2. The number of distinct histories is $n!m!$ in this case.}}
	\label{fig:free+QCD}
\end{figure}

\vspace{-.2cm}
\noindent
The first few  orders of trace-connected diagrams are shown  in   Fig.~\ref{fig:free+QCD}a.  There are $(n{-}1)!$ trace-connected histories containing $n$ incoming/outgoing particles, which correspond to   the cyclic permutations of $n$ elements. All these histories have the same free-theory amplitude. We are thus left with 
\bea
F_0(\beta)=-\beta^{-1}\sum_{n=1}^{\infty} \frac{1}{n} \left(\prod_{i=1}^n  \int\dd^dk_i e^{-\beta E_i} \right)\delta^{(d)}(\bk_1-\bk_2)\delta^{(d)}(\bk_2-\bk_3)\ldots\delta^{(d)}(\bk_n-\bk_1)\,. 
\eea
It is straightforward to integrate the last equation, leading to the   familiar result for the free-theory thermal partition function $F_0(\beta)=\beta^{-1} V_{(d)}\int \dd^dk \ln(1-e^{-\beta E_{\bk}})/(2\pi)^d$. The factor of space volume $V_{(d)}$ comes from a Dirac-delta $\delta^{(d)}(0)$. The presence of one and only one overall spatial delta-distribution is a direct consequence of the requirement of trace-connectedness, and the proportionality to $V_{(d)}$ shows that we have correctly isolated the extensive part of the partition function. Going from bosons (as presented above) to fermions is a matter of signs, which are contained in the normalization of multi-particle states, but the basic logic is the same.

Next we compute  the leading correction to the partition function due to the QCD interactions, i.e. we compute the $O(\alpha_s)$ corrections.  
The   expansion of the logarithm in \eq{master} can be truncated at ${\cal O}(T)$, because each $T\propto\alpha_s$ in that it involves at least one interaction. 
There are infinitely many trace-connected histories one has to sum over, as illustrated in Fig.~\ref{fig:free+QCD}b. 

The interpretation of the diagrams goes as follows. 
We imagine a bunch of particles at past infinity, $n$ of type $I$ and $m$ of type $J$ (this could represent flavor, color, charge, etc.), and study all the trace-connected ways in which this bunch can evolve into itself at future infinity. 
For \emph{ distinguishable }particles free propagation cannot give a trace-connected history and one needs at least a non-trivial scattering $IJ\to IJ$, whose contribution is weighted by the transition amplitude ${M}_{IJ\to IJ}$. We follow the usual convention with $S_{\beta\alpha}=\delta_{\alpha\beta}+(2\pi)^{d{+}1}\delta^{(d{+}1)}(p_\alpha-p_\beta)\,i{M}_{\alpha\to\beta}$.

A single interaction is enough 
to give rise to trace-connected histories for arbitrary $n$ and $m$. Identical particles can exchange among themselves pretty much like for the free propagation, albeit with a combinatorial difference. The factor of $(n!m!)^{-1}$ coming from the trace in \reef{traceO} is in this case completely reabsorbed, as shown by the second and third diagrams contributing to Fig.~\ref{fig:free+QCD}b, where $n=1$ and $m=2$: the two histories are clearly different and should both be counted. With $n=1$ and $m=3$ there are 6 distinct histories, and so on.
The contributions coming from all these histories can be resummed and one finds~\cite{Dashen:1969ep}
\be\label{2to2}
\Delta f=-\frac{1}{2}\,\sum_{I,J} \int \frac{\dd^d k_I}{2E_I(2\pi)^d}\int\frac{\dd^d k_J}{2E_J(2\pi)^d}\,n(E_I)\,n(E_J)\,{M}_{IJ\to IJ}(\bk_I,\bk_J)\,,
\ee
where the amplitude ${M}$ is evaluated in the forward limit, and $n(\omega)=(e^{\beta\omega}\pm 1)^{-1}$ for respectively fermions and bosons, while $\Delta f=(F-F_0)/V_{(d)}$. The factor of $1/2$ is to avoid overcounting. In   ${M}_{IJ\to IJ}(\bk_I,\bk_J)$  the initial and final state kinematic variables are the same. We see that the sum over trace-connected histories is equivalent to a thermal average of (forward) amplitudes.

Let us give here some more details of the derivation of \reef{2to2}, especially how to go from the energy dependent operator $S(E)$ of \reef{master} to ordinary amplitudes. 
Taking the leading term of the expansion in Eq.~\reef{pertT}, we get
\begin{align}\label{FofT}
\Delta F= \int_0^\infty \dd E e^{-\beta E} \int \dd \alpha \,\delta(E-E_\alpha )T_{\alpha\alpha}(E)\,,
\end{align}
where $\int \dd\alpha$ stands for a sum over a complete set of states, as dictated by the trace (the trace-connectedness condition is not crucial for this discussion). {Assuming that $T_{\alpha\alpha}(E)$ is well defined on the support of the Dirac delta}, we can integrate over the latter, finding an expression in terms of the ordinary $T$-matrix element $T_{\alpha\alpha}$ instead of  $T_{\alpha\alpha}(E)$. Since $T_{\beta\alpha}=-(2\pi)^d\delta^{(d)}(\bk_\alpha-\bk_\beta)M_{\alpha\to\beta}$, in the forward limit we get a space-volume factor $(2\pi)^d\delta^{(d)}(0)$. In terms of the free energy density we then find
\be\label{fofM}
\Delta f = -\int \dd\alpha \,e^{-\beta E_\alpha} M_{\alpha\to\alpha}\,.
\ee
Getting to \reef{2to2} is then a just matter of summing over the infinitely many trace-connected $M_{\alpha\to\alpha}$ contributions having only free propagating $I$ and $J$ particles and a single connected amplitude $M_{IJ\to IJ}$, as in Fig.~\ref{fig:free+QCD}b. 

\eq{2to2} instructs us to sum over each pair of particles in the spectrum. In QCD we have 8 colors of gluons coming with helicity + or $-$, plus 6 flavors of quarks, each coming with 3 colors, helicity $\pm1/2$ and chirality $L$ or $R$.


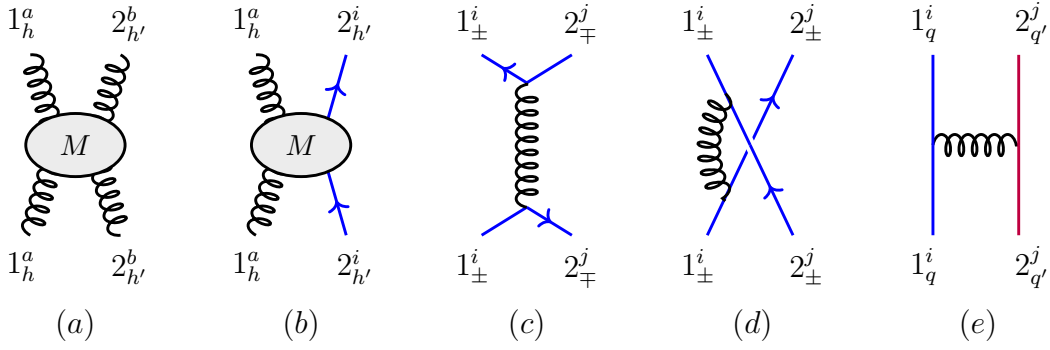
\begin{figure}[t]
\centering
\begin{tikzpicture}[line width=1.1 pt, scale=1.2, baseline=(current bounding box.center)]

	\draw[gluon] (-.5,-1) -- (-.2,0) ;
	\draw[gluon] (-.5,1) -- (-.2,0) ;
	\draw[gluon] (.5,-1) -- (.2,0) ;
	\draw[gluon] (.5,1) -- (.2,0) ;
	\draw[fill=gray!15] (0,0) ellipse (.55 cm and .35 cm) ;
	\node at (0,0) {\small ${M}$} ;
	\node at (-.6,-1.35) {$1^a_h$} ;
	\node at (-.6,1.35) {$1^a_h$} ;
	\node at (.6,-1.35) {$2^b_{h'}$} ;	
	\node at (.6,1.35) {$2^b_{h'}$} ;
	\node at (0,-2) {($a$)} ;

	\draw[gluon] (2.,-1) -- (2.3,0) ;
	\draw[gluon] (2.,1) -- (2.3,0) ;
	\draw[fermion,blue] (3.,-1) -- (2.8,-.3) ;
	\draw[fermion,blue] (2.8,.3) -- (3.,1) ;
	\draw[fill=gray!15] (2.5,0) ellipse (.55 cm and .35 cm) ;
	\node at (2.5,0) {\small ${M}$} ;
	\node at (1.9,-1.35) {$1^a_h$} ;
	\node at (1.9,1.35) {$1^a_h$} ;
	\node at (3.1,-1.35) {$2^i_{h'}$} ;	
	\node at (3.1,1.35) {$2^i_{h'}$} ;
	\node at (2.5,-2) {($b$)} ;

	\draw[blue] (4.5,-1) -- (5,-.69) ;
	\draw[fermion,blue]  (5,-.69) -- (5.5,-1) ;
	\draw[fermion,blue]  (5,.69) -- (4.5,1) ;
	\draw[blue] (5.5,1) -- (5,.69) ;
	\draw[gluon] (5,-.69) -- (5,.69) ;
	\node at (4.4,-1.35) {$1^i_{\pm}$} ;
	\node at (4.4,1.35) {$1^i_{\pm}$} ;
	\node at (5.6,-1.35) {$2^j_{\mp}$} ;	
	\node at (5.6,1.35) {$2^j_{\mp}$} ;
	\node at (5,-2) {($c$)} ;

	\draw[blue] (7,-1) -- (7.475,0) ;
	\draw[fermion,blue] (7.475,0) -- (7.95,1) ;
	\draw[white,line width =4] (7.95,-1) -- (7.0,1) ;
	\draw[fermion,blue] (7.95,-1) -- (7.475,0) ;
	\draw[blue] (7.475,0) -- (7.0,1) ;
	\draw[gluon] (7.2,.57) to [bend right] (7.2,-.62) ;
	\node at (6.9,-1.35) {$1^i_\pm$} ;
	\node at (6.9,1.35) {$1^i_\pm$} ;
	\node at (8.1,-1.35) {$2^j_\pm$} ;	
	\node at (8.1,1.35) {$2^j_\pm$} ;
	\node at (7.5,-2) {($d$)} ;

	\draw[gluon] (9.5,0) -- (10.45,0) ;
	\draw[blue] (9.5,-1) -- (9.5,0) ;
	\draw[blue] (9.5,0) -- (9.5,1) ;
	\draw[purple] (10.45,0) -- (10.45,-1) ;
	\draw[purple] (10.45,0) -- (10.45,1) ;
	\node at (9.4,-1.35) {$1^i_{q}$} ;
	\node at (9.4,1.35) {$1^i_{q}$} ;
	\node at (10.6,-1.35) {$2^j_{q'}$} ;	
	\node at (10.6,1.35) {$2^j_{q'}$} ;
	\node at (10,-2) {($e$)} ;
					
 \end{tikzpicture}
 \caption{\emph{Forward diagrams entering the computation of ${\cal O}(g_s^2)$ thermal effects in QCD. Notice that diagrams (d) and (e) are topologically distinct in that quarks form respectively one and two loops, due to a different initial and final state identification. Diagram (e) is the only one allowing independent quark flavors, but it vanishes because $\tr \,t^a=0$.}}
 \label{fig:QCD}
 \end{figure}

Let us start from the pure gluon sector, where the relevant amplitude in the all-incoming convention is given by the celebrated Parke-Taylor formula
\be\label{gluon_ampli}
{M}(1_-^a,2_-^b,3_+^c,4_+^d)=-2g_s^2\s{12}^4\left[\frac{f^{abe}f^{cde}}{\s{12}\s{23}\s{34}\s{41}}+\frac{f^{ace}f^{bde}}{\s{13}\s{32}\s{24}\s{41}}\right],
\ee
written using the spinor-helicity formalism (for textbook expositions see for instance \cite{Henn:2014yza,Elvang:2015rqa}). From this amplitude we can extract, by appropriately crossing, the in-out amplitudes for the three non-vanishing helicity configurations $\pm\pm\to\pm\pm$ and $+-\to +-$.

In order to take the trace we need to evaluate the forward limit of the gluon amplitude  in \reef{gluon_ampli}. Taking for example the channel $1,2\to 3,4$, with the trace identification defined by $3\equiv 1,4\equiv 2$, we see that the first term is non-singular and gives $-2g_s^2\sum_c f^{abc}f^{abc}$, while the second one has a $0/0$ ambiguity that has to be cured. If we go off the forward limit, we see that the second term of \reef{gluon_ampli} is zero if we identify the colors of the initial and final states as in Fig.~\ref{fig:QCD}$a$, because it is proportional to $\sum_c f^{aac}f^{bbc}{=}0$. Therefore the $0/0$ limit should be understood as $\lim_{t\to 0}{0}/{t}=0$, and the potentially dangerous term of \reef{gluon_ampli} actually gives zero when inserted in \reef{2to2}.

Plugging the forward gluon amplitude into \reef{2to2} and using that $\int_0^\infty dk k/(e^k-1)=\pi^2/6$, after summing over color and helicity we get
\be
f_{\rm gluons}=\alpha_s N_{\rm c} (N_{\rm c}^2-1)\frac{\pi T^4}{36}\,,
\ee
where $\alpha_s=g_s^2/4\pi$. 
To sum over colors we  have used $f^{abc}f^{abc}=N_{\rm c} (N_{\rm c}^2-1)$. Note that the $+-$ and $-+$ helicity configurations are to be treated as distinct contributions to \reef{2to2}.

Let us now consider the effect of quarks. 
We start with   the contribution of a single flavor, say $q=u_L$, coming with 2 helicities and $N_{\rm c}=3$ colors (we treat $u_R$ as a different flavor).
We need to take  into account all possible $2\to 2$ forward scatterings  involving at least one quark. A gluon $g$ of any helicity and color can encounter a quark $q$ with any helicity and color, and the same for a quark-quark encounter. We are left to evaluate
\be
f_{q}=-\sum_{\substack{a,i,h,h'}}\int \dd^3_{\rm B} k\,\dd^3_{\rm F} k'~{M}\!\left(g_h^a(\bk),q_{h'}^{i}({\bk'})\right)-\frac{1}{2}\sum_{\substack{i,j,h,h'}}\int \dd^3_{\rm F} k\,\dd^3_{\rm F} k'~{M}\!\left(q_{h}^{i}({\bk}),q_{h'}^{j}({\bk'})\right)\,,
\ee
where we introduced the single-particle thermal measures $\dd^d_{\rm F} k$ and $\dd^d_{\rm B} k$ for fermions and bosons, given respectively by $\dd^d k /2E(2\pi)^d(e^{\beta E}\pm 1)$. 
The forward amplitude is simply a numerical coefficient and the $t\to 0$ singularities, present in every channel, also in this case delicately cancel thanks to the color structure in the forward configuration. After summing over colors we get 
 that $\sum_{a,i}{M}(g^a,q^i)=\sum_{i,j}{M}(q^i,q^j)=-g_s^2 (N_{\rm c}^2-1)$ for both terms, and therefore 
\be\label{Deltaf_quarks}
f_{q}=4\,g_s^2 (N_{\rm c}^2-1) \bigg[ \int\dd^3_{\rm B} k\int \dd^3_{\rm F} k+\frac{1}{2}\left(\int \dd^3_{\rm F} k\right)^2\,\bigg]=\alpha_s (N_{\rm c}^2-1)\frac{5\pi T^4}{288}\,,
\ee
where the factor of 4 comes from the independent helicity choices.

When we consider more flavors and account for both chiralities $L$ and $R$, we see that the only new effect comes from diagrams like the one in Fig.~\ref{fig:QCD}$e$. However the sum over colored gluons enforces the vanishing of the amplitude when initial and final state quantum numbers are identified (as dictated by the trace). All in all this implies  that,   in order to take into account more flavors, one needs just to multiply \eq{Deltaf_quarks} by $2N_{\rm f}$ (the 2 is for $L$ and $R$). We find
\be
 f_{\rm QCD}=\alpha_s (N_{\rm c}^2-1)\,\frac{\pi T^4}{36}\left(N_{\rm c}+\frac{5}{4}N_{\rm f}\,\right) \, , \label{fqcd}
\ee
in perfect agreement with textbook thermal field theory calculations --- see e.g. chapter~5.4 in \cite{Laine:2016hma} and references therein ---, which involve loops of ghost-fields and gauge fixing terms. The two-loop $O(\alpha_s)$ result  \reef{fqcd}   is  obtained by  tracing over the two-to-two tree-level matrix elements.  
 This calculation  makes apparent that only thermal (non-vacuum) loops are relevant at this order.

With this example in mind we can make a first important comment on the qualitative structure of Eq.~\reef{master}. The master formula instructs us that, to extract the free energy of a given system, we need to take a thermally weighted average of the relevant scattering amplitudes among the constituents of the system. The scattering amplitudes  are well-defined distributions of the external kinematics when this is \emph{generic} (that is, $S_{\beta\alpha}$ is a distribution in $\dd \alpha\,\dd \beta$). However in the trace formula amplitudes are evaluated in the forward limit $\beta\equiv\alpha$ and are not guaranteed to be well-defined distributions anymore. This problem was circumvented in the previous analysis because the potentially dangerous contributions vanished, but it becomes ubiquitous when considering NL effects. We will see in several examples in the paper that {\bf(a)} the sick behavior is due to a subset of diagrams with precisely traceable kinematic features and that {\bf(b)} smearing the $S$-matrix elements in the energy variable $E$, as in \reef{master}, is crucial in order to resolve these singularities, which are of IR nature, by making $S_{\alpha\alpha}(E)$ a well-defined distribution in $\dd E\,\dd \alpha$.

To study the structure of higher order effects to the free energy, as organized by \reef{master}, we consider the theory of a long Flux Tube, both because all-order results are available and because it has no physical   IR divergences -- i.e.  IR divergences encountered in  intermediate steps of the calculation are guaranteed to cancel out if properly handled.

\section{Flux Tube free-energy}
\label{sec:fluxtuberecap}

\subsection{Flux Tube recap}

At long distances   a long Flux Tube is described by an Effective Field Theory  given by the Nambu-Goto (NG) action perturbed by a series of curvature invariants -- see refs.~\cite{Luscher:2004ib,Teper:2009uf,Dubovsky:2012sh,Aharony:2013ipa} and references therein. 
The  effective action is given by
$
A=\int d^2\sigma \sqrt{-h} (\ell_s^{-2}+\ldots) \, , 
$
where $h$ is the determinant of the induced metric $h_{\alpha\beta}=\partial_\alpha X^\mu \partial_\beta X^\nu \eta_{\mu\nu}$, 
 $\eta$ is the metric on the $D$-dimensional embedding space-time and $X^\mu(\sigma)$ are the embedding coordinates of the Flux Tube world-sheet in spacetime. 
 
The dots $\ldots$ in  the action $A$ involve higher-dimensional local operators built out of the induced metric and the covariant derivatives. 
Interestingly, the first operators with non-vanishing effects appear at $\order{\partial^8 X^4}\sim \order{K^4}$ where $K^\mu_{\alpha\beta}\equiv \nabla_\alpha \partial_\beta X^\mu$ is the extrinsic curvature tensor.
These (and other) higher dimensional operators are introduced in an EFT fashion to parametrize deviations from the pure NG theory needed to realistically model e.g. a Yang Mills Flux Tube, as apparent from lattice simulations.
We will see the effect on the free energy of the lowest higher-dimensional invariant in Section~\ref{sec:nonuniversal}.

 In the static gauge,  $X^\mu(\sigma )=(\sigma^\alpha, X^a(\sigma))$ where $a=1,\dots , D-2$ are the transverse excitations of the Flux Tube string. 
The scattering amplitude of the transverse excitations $X^a$ is an interesting observable with a rich phenomenology~\cite{Dubovsky:2013gi,Conkey:2016qju,Dubovsky:2015zey,EliasMiro:2019kyf,EliasMiro:2021nul,Gaikwad:2023hof}.
The $D-2$ transverse excitations are massless Nambu-Goldstone bosons that transform as vectors of the $O(D-2)$ symmetry. 
For simplicity we first consider the Scattering matrix of  a single degree of freedom, i.e. $D=3$ target spacetime, and then generalize to  $D>3$.
For our current purposes the only bits of information that we need about $D=3$ Flux Tubes are the following.

There is a realization of the $D=3$ flux tube that is exactly solvable in the integrable $S$-matrix sense:
there is no particle production $M_{n\to m> n}=0$, the $2\to 2$ $S$-matrix is exactly known, and the $n\to n$ $S$-matrix elements factorize into products of $2\to 2$. 
The $S$-matrix is given by
 \be
\langle \,q_1,q_2,\ldots,q_n \,|\,\mathbb{S}\, |\, p_1,p_2,\ldots,p_n\,\rangle= e^{\frac{i \ell_s^2}{4}\sum_{i<j}^n s_{ij}  }   \,  \langle \,q_1,q_2,\ldots,q_n \,|\, p_1,p_2,\ldots,p_n\,\rangle
\label{Sexact}
\ee
where $s_{ij}=2p_i \cdot p_j$, and note that unitarity is satisfied $\mathbb{S}^\dagger \mathbb{S}=\mathbb{1}$.

Besides  the Scattering amplitude of long Flux Tube excitations, another   interesting  observable is   the   finite temperature  free-energy --- often called   the Casimir energy --- of a closed Flux Tube of length  $\beta $.~\footnote{This observable can be determined in Lattice Monte Carlo simulations  of Yang-Mills theory by studying the decay of the correlation function of two Polyakov loops of length $\beta$ --- see \cite{Teper:2009uf} for a review.}
For the $D=3$ integrable realization of the Flux Tube theory, the free energy density  is given by
 \be
f(\beta) = \frac{1}{\ell_s^2}\, \sqrt{1-\frac{\ell_s^2}{\beta^2} \, \frac{\pi}{3}}  =\frac{1}{\ell_s^2}-\frac{\pi}{6\beta^2} - \frac{\ell_s^2\pi^2}{72\beta^4} - \frac{\ell_s^4\pi^3}{432\beta^6}- \frac{5\ell_s^6\pi^4}{10368\beta^8}+\ldots  \label{Esqrt} 
\ee
where $f(\beta)=\beta^{-1} E_0(\beta)$, with $E_0$  the Casimir or vacuum energy. Eq.~\reef{Esqrt} has been established in ref.~\cite{Dubovsky:2012wk} using the TBA. The integrable $D=3$ flux tube is a   precious example   to test  \reef{masterformula}: both 
the  $S$-matrix  \reef{Sexact} and free energy \reef{Esqrt}  are known exactly.

We will need to input $M$-matrix elements into \reef{master}. 
These are obtained from the interacting part of $\mathbb{S}$ in \reef{Sexact} by stripping a $i(2\pi)^2\delta^{(2)}(p_{\rm in}-p_{\rm out})$, and can be expressed as
\be\label{Mnm}
\begin{tikzpicture}[line width=1.1 pt, scale=.33, baseline=(current bounding box.center)]

		\draw[cyan] (6,0) -- (0,6) ;
		\draw[cyan] (7,1) -- (1,7) ;
		\draw[cyan] (9,3) -- (3,9) ;
		\draw[purple] (3,0) -- (9,6) ;
		\draw[purple] (2,1) -- (8,7) ;
		\draw[purple] (0,3) -- (6,9) ;

		\draw[black,fill=gray!15] (4.5,1.5) circle (.35 cm) ;
		\draw[black,fill=gray!15] (4.5,3.5) circle (.35 cm) ;
		\draw[black,fill=gray!15] (3.5,2.5) circle (.35 cm) ;
		\draw[black,fill=gray!15] (5.5,2.5) circle (.35 cm) ;
		\draw[black,fill=gray!15] (1.5,4.5) circle (.35 cm) ;
		\draw[black,fill=gray!15] (2.5,5.5) circle (.35 cm) ;
		\draw[black,fill=gray!15] (7.5,4.5) circle (.35 cm) ;
		\draw[black,fill=gray!15] (6.5,5.5) circle (.35 cm) ;
		\draw[black,fill=gray!15] (4.5,7.5) circle (.35 cm) ;
		
		\draw[cyan,dotted,line width=1.5] (4.13,5.13) -- (4.87,5.87) ;
		\draw[purple,dotted,line width=1.5] (2.17,2.43) -- (1.43,3.17) ;
		
		\node at (6.4,-.5) {\footnotesize 1} ;
		\node at (7.4,.5) {\footnotesize 2} ;
		\node at (9.4,2.5) {\footnotesize $m$} ;
		\node at (2.7,-.7) {\scriptsize $m{+}1$} ;
		\node at (1.7,.3) {\scriptsize $m{+}2$} ;
		\node at (-.3,2.3) {\scriptsize $m{+}n$} ;

\end{tikzpicture}
= -i\,\frac{ J_{m+n}}{2\pi^2} \left( e^{\frac{i \ell_s^2}{4}\sum_{i<j}^{m+n} s_{ij}  } -1\right) \sum_{\sigma_l,\sigma_r} \delta^{(m+n-2)}_{\sigma_l,\sigma_r}(E_{\rm in},E_{\rm out})\,,
\ee
where $J_k=\prod_{i=1}^k 4\pi E_i$, and
$
\delta^{(m+n-2)}_{\sigma_l,\sigma_r}(E_{\rm in},E_{\rm out}) \equiv  \prod_{i=1}^{m-1}\delta\left(E_i-E'_{\sigma_l(i)}\right)\prod_{j=m+1}^{m+n-1}\delta\left(E_j-E'_{\sigma_r(j)}\right)
$, 
with $\sigma_l$ and $\sigma_r$ denoting respectively the $m!$ and $n!$ permutations acting on the massless left- ($l$-) and right- ($r$-)moving outgoing particles.


\subsection{Leading Order correction}\label{sec:LO}

The aim of this and the following sections is to illustrate how to compute the free-energy \reef{Esqrt} via 
the DMB formulation  \reef{master} using the integrable $S$-matrix \reef{Sexact} at $D=3$ as input. The bigger goal is to gain insight of the trace formula, both in its general aspects and in its specific application to the Flux Tube theory.

The first order of business is to discuss the   $O(\ell_s^2)$  effects, which are captured by \reef{2to2} and require knowledge of the $2{\to}2$ on-shell amplitude in the forward limit, given by
\be\label{flux2to2}
M(s)= \frac{\ell_s^2s^2}{2}+\frac{i \ell_s^4 s^3}{16}+\cdots.
\ee
This contribution to the free energy of the Flux Tube is depicted in diagram (a) of Fig.~\ref{fig:tubeofflux}. 

The $M$ matrix   is only non-zero when the two particles move in opposite directions, so they can be treated as distinguishable and we can remove the 1/2! from \reef{2to2}, to find 
\bea\label{deltaF2flux}
 f_{\rm LO}=-\int_0^\infty \frac{\dd E_1n_1}{4\pi E_1}  \int_0^\infty\frac{\dd E_2n_2}{4\pi E_2} \ 8 \ell_s^2 E_1^2E_2^2 =-\frac{\ell_s^2\pi^2}{72\beta^4} \ , 
\eea
where we denoted $n_i=n(E_i)$. The result nicely matches the  $O(\ell_s^2)$ term  of  \reef{Esqrt}. In the following, we will make abundant use of diagrammatic representations. For example, the contribution to the free energy we just computed could be denoted in either of the following ways
\be
 \begin{tikzpicture}[line width=1.1 pt, scale=.6, baseline=(current bounding box.center)]
	
		\draw[cyan] (0,0) circle (.65cm) ;
		\draw[purple] (1.3,0) circle (.65cm) ;
		\draw[black,fill=gray!15] (0.65,0) circle (.1 cm) ;
		
		\begin{scope}[shift={(4,0)}]
		\draw[purple] (-.5,-.5) -- (.5,.5) ;
		\draw[cyan] (.5,-.5) -- (-.5,.5) ;
		\draw[black,fill=gray!15] (0,0) circle (.1 cm) ;
		\node at (-.55,-.75) {\tiny 1} ;
		\node at (-.55,.75) {\tiny 2} ;
		\node at (.55,-.75) {\tiny 2} ;
		\node at (.55,.75) {\tiny 1} ;
		\end{scope}
		
		\node at (-2.2,0) {$f_{\rm LO}~=$} ;
		\node at (2.7,0) {=} ;
		
	 \end{tikzpicture}
\ee
with $l$-moving particles denoted in cyan and $r$-moving particles in purple. The structure of the $S$-matrix is such that each interaction is a 4-vertex involving 2 $l$- and 2 $r$-movers. The first representation emphasizes the topological structure of the trace-connected diagram, while the second one emphasizes what amplitude we are tracing. As we will see, at the next orders the meaning of the two representations will not coincide anymore, as the first kind will encompass a larger set of objects than the second.

\subsection{NLO correction: generalities and regular contribution}\label{sec:ls4reg}

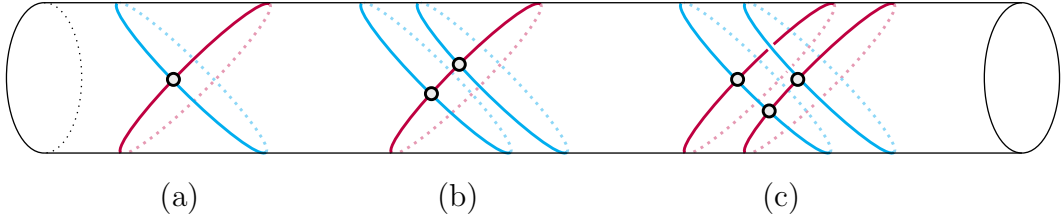
\begin{figure}[t]
\centering
\begin{tikzpicture}[line width=1.1 pt, scale=1.0, baseline=(current bounding box.center)]

	\begin{scope}[shift={(-3.5,0)}]
	\begin{scope}[shift={(0,-0.02)}]
 	\draw[purple,rotate around={45:(0,1)}] (0,1) arc(0:180:1.4cm and .2cm);
	\draw[purple!40,dotted,rotate around={45:(0,1)}] (0,1) arc(0:160:1.4cm and -.2cm);
	\end{scope};
	
	\begin{scope}[shift={(-0.05,0.01)}]
 	\draw[cyan!40,dotted,rotate around={-45:(0,-1)}] (0,-1) arc(0:170:1.4cm and .2cm);
	\draw[cyan,rotate around={-45:(0,-1)}] (0,-1) arc(0:180:1.4cm and -.2cm);
	\end{scope};
	
	\draw[black,fill=gray!15] (-1.28,-.02) ellipse (.08 cm and .08 cm) ;
	\end{scope}
	
	
	\begin{scope}[shift={(.5,0)}]
	
	\begin{scope}[shift={(-.4,-0.02)}]
	\draw[purple!40,dotted,rotate around={45:(0,1)}] (0,1) arc(0:160:1.4cm and -.2cm);
	\end{scope};
	
	\begin{scope}[shift={(-0.8,0.01)}]
 	\draw[cyan!40,dotted,rotate around={-45:(0,-1)}] (0,-1) arc(0:170:1.4cm and .2cm);
	\draw[cyan,rotate around={-45:(0,-1)}] (0,-1) arc(0:180:1.4cm and -.2cm);
	\end{scope};
	
	\begin{scope}[shift={(-0.05,0.01)}]
 	\draw[cyan!40,dotted,rotate around={-45:(0,-1)}] (0,-1) arc(0:170:1.4cm and .2cm);
	\draw[cyan,rotate around={-45:(0,-1)}] (0,-1) arc(0:180:1.4cm and -.2cm);
	\end{scope};
	
	\begin{scope}[shift={(-.4,-0.02)}]
 	\draw[purple,rotate around={45:(0,1)}] (0,1) arc(0:180:1.4cm and .2cm);
	\end{scope};
	
	\draw[black,fill=gray!15] (-1.85,-.21) ellipse (.08 cm and .08 cm) ;
	\draw[black,fill=gray!15] (-1.48,.18) ellipse (.08 cm and .08 cm) ;
	\end{scope}
	
	
	\begin{scope}[shift={(4,0)}]
	
	\begin{scope}[shift={(0,-0.02)}]
	\draw[purple!40,dotted,rotate around={45:(0,1)}] (0,1) arc(0:160:1.4cm and -.2cm);
	\end{scope};
	
	\begin{scope}[shift={(.8,-0.02)}]
	\draw[purple!40,dotted,rotate around={45:(0,1)}] (0,1) arc(0:160:1.4cm and -.2cm);
	\end{scope};
	
	\begin{scope}[shift={(-0.05,0.01)}]
 	\draw[cyan!40,dotted,rotate around={-45:(0,-1)}] (0,-1) arc(0:170:1.4cm and .2cm);
	\draw[cyan,rotate around={-45:(0,-1)}] (0,-1) arc(0:180:1.4cm and -.2cm);
	\end{scope};
	
	\begin{scope}[shift={(.8,0.01)}]
 	\draw[cyan!40,dotted,rotate around={-45:(0,-1)}] (0,-1) arc(0:170:1.4cm and .2cm);
	\end{scope};
	
	\begin{scope}[shift={(0,-0.02)}]
 	\draw[purple,rotate around={45:(0,1)}] (0,1) arc(0:180:1.4cm and .2cm);
	\end{scope};
	
	\begin{scope}[shift={(.8,-0.02)}]
 	\draw[purple,rotate around={45:(0,1)}] (0,1) arc(0:180:1.4cm and .2cm);
	\end{scope};
	
	\begin{scope}[shift={(.8,0.01)}]
	\draw[white,line width=3,rotate around={-45:(0,-1)}] (0,-1) arc(0:180:1.4cm and -.2cm);
	\draw[cyan,rotate around={-45:(0,-1)}] (0,-1) arc(0:180:1.4cm and -.2cm);
	\end{scope};
	
	\draw[black,fill=gray!15] (-1.28,-.02) ellipse (.08 cm and .08 cm) ;
	\draw[black,fill=gray!15] (-1.28+.8,-.02) ellipse (.08 cm and .08 cm) ;
	\draw[black,fill=gray!15] (-1.28+.42,-.02-.42) ellipse (.08 cm and .08 cm) ;
	
	\end{scope}
	
	
	\draw[line width = .5] (-6.5,1) arc(90:270:.5cm and 1cm);
	\draw[line width = .5,black!100,dotted] (-6.5,1) arc(90:270:-.5cm and 1cm);
	\draw[line width = .5] (6.5,1) arc(90:270:.5cm and 1cm);
	\draw[line width = .5] (6.5,1) arc(90:270:-.5cm and 1cm);
	
	\draw[line width = .5] (-6.5,1) -- (6.5,1) ;
	\draw[line width = .5] (-6.5,-1) -- (6.5,-1) ;
	
	\node at (-4.7,-1.6) {(a)} ;
	\node at (-1,-1.6) {(b)} ;
	\node at (3.3,-1.6) {(c)} ;

 \end{tikzpicture}
 \caption{\emph{Forward diagrams entering the computation of the free energy of the flux tube. Moving from left to right, they contribute at order $\ell_s^2$, $\ell_s^4$ and $\ell_s^6$.}}
 \label{fig:tubeofflux}
 \end{figure}

Contrary to the LO correction, the NLO contribution to the DMB formula \reef{master} shows a formidable complexity, and before going into its computation we would like to pave the way with a number of organizing principles and general comments.
\begin{itemize}
\item[1.] The first rationale for organizing the various contributions is topology. Each term in the DMB formula can be unambiguously associated to a vacuum graph, each node corresponding to an interaction vertex. For the case at hand, with only quartic interactions involving a left- and right-mover due to the underlying integrability, we have at NLO
	\begin{equation}\label{vac34}
	\begin{tikzpicture}[line width=1.1 pt, scale=.7, baseline=(current bounding box.center)]
	
		\draw[cyan] (0,0) ellipse (.8cm and .4cm) ;
		\draw[purple] (0,0) circle (.8cm) ;
		\draw[black,fill=gray!15] (0.77,0) circle (.1 cm) ;
		\draw[black,fill=gray!15] (-0.77,0) circle (.1 cm) ;
		\node at (1.2,0) {\scriptsize $\ell_s^2$} ;
		\node at (-1.2,0) {\scriptsize $\ell_s^2$} ;
		
		\node at (3.3,0) {\&} ;
		
		\begin{scope}[shift={(1,0)}]
		\draw[cyan] (4.8,0) circle (.6cm) ;
		\draw[cyan] (7.2,0) circle (.6cm) ;
		\draw[purple] (6,0) circle (.6cm) ;
		\draw[black,fill=gray!15] (5.4,0) circle (.1 cm) ;
		\draw[black,fill=gray!15] (6.6,0) circle (.1 cm) ;
		\node at (5.,0) {\scriptsize $\ell_s^2$} ;
		\node at (7,0) {\scriptsize $\ell_s^2$} ;
		\end{scope}
		
	 \end{tikzpicture}~~,
	 \end{equation}
where the first (melon) and second (caterpillar) graph correspond to respectively a regular and a singular contribution to the trace formula in the forward limit. To the second graph, which should obviously be complemented with a similar graph with $l$ and $r$ swapped, belongs diagram (b) of Fig.~\ref{fig:tubeofflux}. Surprisingly, the two contributions will undergo a somehow different treatment. Notice that on top of these new topologies there is also
	\be\label{vac24}
	\begin{tikzpicture}[line width=1.1 pt, scale=.7, baseline=(current bounding box.center)]
	
		\draw[cyan] (0,0) circle (.65cm) ;
		\draw[purple] (1.3,0) circle (.65cm) ;
		\draw[black,fill=gray!15] (0.65,0) circle (.1 cm) ;
		\node at (1.1,0) {\scriptsize $\ell_s^4$} ;
		
	 \end{tikzpicture}~~,
	 \ee
which has the same topology of the LO contribution, whereas now the interaction vertex comes from the $O(\ell_s^4)$ contribution to \reef{flux2to2}, which is a one-loop effect.
\item[2.] Contributions to \reef{master} can also be organized according to the number of $T$-matrix insertions (cf. \reef{pertT}). To include all $O(\ell_s^4)$ effects we need to consider up to two $T$ insertions.
\item[3.] At a given order in $T$ a further classification can be made, based on the number of legs of the relevant $T$-matrix element or on the topology of the diagram. At $O(T)$ there are three topologically distinct contributions corresponding to connected $T$ insertions
\be
	\begin{tikzpicture}[line width=1.1 pt, scale=.7, baseline=(current bounding box.center)]
	
		\draw[purple] (-.5,-.5) -- (.5,.5) ;
		\draw[cyan] (.5,-.5) -- (-.5,.5) ;
		\draw[black,fill=gray!15] (0,0) circle (.1 cm) ;
		\node at (0.6,0) {\scriptsize $\ell_s^4$} ;
		
		\begin{scope}[shift={(4,0)}]
		\draw[purple] (-.6,-.6) -- (.6,.6) ;
		\draw[cyan] (.3,-.7) -- (-.7,.3) ;
		\draw[cyan] (.7,-.3) -- (-.3,.7) ;
		\draw[black,fill=gray!15] (-.2,-.2) circle (.1 cm) ;
		\draw[black,fill=gray!15] (.2,.2) circle (.1 cm) ;
		\node at (.5,-.9) {\scriptsize 1} ;
		\node at (.9,-.5) {\scriptsize 2} ;
		\node at (-.9,.5) {\scriptsize 1} ;
		\node at (-.5,.9) {\scriptsize 2} ;
		\end{scope}
		
		\begin{scope}[shift={(8,0)}]
		\draw[purple] (-.6,-.6) -- (.6,.6) ;
		\draw[cyan] (.3,-.7) -- (-.7,.3) ;
		\draw[cyan] (.7,-.3) -- (-.3,.7) ;
		\draw[black,fill=gray!15] (-.2,-.2) circle (.1 cm) ;
		\draw[black,fill=gray!15] (.2,.2) circle (.1 cm) ;
		\node at (.5,-.9) {\scriptsize 1} ;
		\node at (.9,-.5) {\scriptsize 2} ;
		\node at (-.9,.5) {\scriptsize 2} ;
		\node at (-.5,.9) {\scriptsize 1} ;
		\end{scope}
		
		\node at (2,0) {,};
		\node at (6,0) {,};
		\node at (10,0) {,};
		
	 \end{tikzpicture}
	 \ee
the last two $T_{3{\to}3}$ diagrams belonging to respectively the `caterpillar' and `melon' topology due to the different trace identifications of the $l$-moving particles.
On top of these three there are two types of diagrams corresponding to \emph{disconnected} single $T$ insertions that become connected only after the trace is taken
\be
	\begin{tikzpicture}[line width=1.1 pt, scale=.7, baseline=(current bounding box.center)]
	
		\draw[cyan] (-.5,-.5) -- (.5,.5) ;
		\draw[purple] (.5,-.5) -- (-.5,.5) ;
		\draw[cyan] (1,-.5) -- (2,.5) ;
		\draw[purple] (2,-.5) -- (1,.5) ;
		\draw[black,fill=gray!15] (0,0) circle (.1 cm) ;
		\draw[black,fill=gray!15] (1.5,0) circle (.1 cm) ;
		\node at (-.5,-.8) {\scriptsize 1} ;
		\node at (.5,-.8) {\scriptsize 2} ;
		\node at (1,-.8) {\scriptsize 3} ;
		\node at (2,-.8) {\scriptsize 4} ;
		\node at (-.5,.8) {\scriptsize 4} ;
		\node at (.5,.8) {\scriptsize 3} ;
		\node at (1,.8) {\scriptsize 2} ;
		\node at (2,.8) {\scriptsize 1} ;
		
		\begin{scope}[shift={(6,0)}]
		\draw[cyan] (-.5,-.5) -- (.5,.5) ;
		\draw[purple] (.5,-.5) -- (-.5,.5) ;
		\draw[cyan] (1,-.5) -- (2,.5) ;
		\draw[purple] (2,-.5) -- (1,.5) ;
		\draw[black,fill=gray!15] (0,0) circle (.1 cm) ;
		\draw[black,fill=gray!15] (1.5,0) circle (.1 cm) ;
		\node at (-.5,-.8) {\scriptsize 1} ;
		\node at (.5,-.8) {\scriptsize 2} ;
		\node at (1,-.8) {\scriptsize 3} ;
		\node at (2,-.8) {\scriptsize 4} ;
		\node at (-.5,.8) {\scriptsize 4} ;
		\node at (.5,.8) {\scriptsize 1} ;
		\node at (1,.8) {\scriptsize 2} ;
		\node at (2,.8) {\scriptsize 3} ;
		\end{scope}
		
		\node at (3.8,0) {\&};
		\node at (9,0) {,};
		
	 \end{tikzpicture}
	 \ee
belonging respectively to the `melon' and `caterpillar' topology. For what concerns this last topology, there is also a $l{-}r$ mirror contribution that we did not depict. To conclude, at $O(T^2)$ there are two more graphs contributing at NLO
\be
	~~~\begin{tikzpicture}[line width=1.1 pt, scale=.65, baseline=(current bounding box.center)]
		
		\draw[cyan] (-.5,-.5) -- (.5,.5) ;
		\draw[purple] (.5,-.5) -- (-.5,.5) ;
		\draw[dashed,line width=.5] (-.9,.6) -- (.9,.6) ;
		\draw[cyan] (.5,.7) -- (-.5,1.7) ;
		\draw[purple] (-.5,.7) -- (.5,1.7) ;
		\draw[black,fill=gray!15] (0,0) circle (.1 cm) ;
		\draw[black,fill=gray!15] (0,1.2) circle (.1 cm) ;
		
		\begin{scope}[shift={(5.5,.5)}]
		\draw[purple] (-.8,-.8) -- (.8,.8) ;
		\draw[cyan] (.2,-.8) -- (-.8,.2) ;
		\draw[cyan] (.8,-.2) -- (-.2,.8) ;
		\draw[black,fill=gray!15] (-.3,-.3) circle (.1 cm) ;
		\draw[black,fill=gray!15] (.3,.3) circle (.1 cm) ;
		\draw[white,line width=2] (-1,0) -- (1,0) ;
		\draw[dashed,line width=.5] (-1.05,0) -- (1.05,0) ;
		\node at (.4,-1) {\scriptsize 1} ;
		\node at (1,-.4) {\scriptsize 2} ;
		\node at (-1,.4) {\scriptsize 1} ;
		\node at (-.4,1) {\scriptsize 2} ;
		\end{scope}
		
		\node at (2.6,0.6) {\&};
		\node at (7.5,0.5) {,};

	 \end{tikzpicture}
	 \ee
together as usual with the $l{-}r$ symmetric version of the second `caterpillar' diagram.
\item[4.] At $O(\ell_s^4)$ the remaining trace-connected contributions to \reef{master}, which are infinitely many, can only differ from the above ones by a number of free propagations, which correspond to windings and whose effects can easily be resummed. The diagrams whose windings are easiest to resum are those with a single $T_{m{\to}m}$ insertion. In this case one gets a total of $m$ Bose-Einstein or Fermi-Dirac densities depending on the statistics of the particles partaking the scattering. For multiple $T$ insertions we get densities also from the intermediate particles, but we need to compensate with a factor of $e^{\beta E_i}$ for each one of them. This is dictated by the master formula \reef{master}. We have in fact an integral $\int \dd E e^{-\beta E}$ times a function that always has support on $E=E_\alpha$, the total energy of the state(s) we use to resolve the trace. If there are internal particles winding, the factor of $e^{-\beta E_\alpha}$ does not include their energy suppression, so in order to reconstruct complete densities we need to compensate with a positive exponential.
\end{itemize}
To both the melon and the caterpillar vacuum diagrams there correspond three distinct on-shell diagrams. This proliferation of diagrams can only worsen as we go to higher orders. However we will see that there is a strong interplay among diagrams belonging to the same topology, and the physical information is contained in only a few, the remaining diagrams being there only to remove unphysical contributions like spurious singularities, imaginary parts, etc.

The $3\to 3$  amplitude can be extracted from \reef{Mnm}, and is given at tree-level and  $O(\ell_s^4)$ by
\be\label{123to456}
M_{1_l2_l3_r\to 4_l5_l6_r}=32 \pi i\,\ell_s^4\,E_1^2E_2^2E_3^3\,\left[\,\delta(E_1-E_4)+\delta(E_1-E_5)\,\right],
\ee
with an analogous result for the mirror symmetric amplitude obtained by exchanging $l$- and $r$-moving particles.
The presence of an extra Dirac delta (besides the delta functions for total two-momentum conservation) is a manifestation of the constraints coming from the underlying integrability of the Flux Tube theory.

Eq.~\reef{123to456}, which can be  computed via  standard Feynman diagram techniques (see appendix~\ref{ls4_appendix}) or right-away from the exact factorized $S$-matrix \reef{Sexact}, is perfectly well defined as a distribution over the 6-particle kinematic space (which is a 4-manifold in 1+1 spacetime dimensions). However, when we restrict to the forward configurations with $E_{i+3}\equiv E_i$, which make up a 3-manifold, the first term of \reef{123to456} becomes singular and makes the amplitude ill-defined.

The singular term corresponds to the caterpillar topology, so we can set it aside for the moment according to the first organizing  rationale, topology, and focus on the regular diagram and its topological companions. Taking the trace and accounting for the windings as explained before, we find an integral over three Bose-Einstein densities ($\beta=1$)
\begin{align}\label{melon33}
\begin{tikzpicture}[line width=1.1 pt, scale=.7, baseline=(current bounding box.center)]
 		\draw[purple] (-.6,-.6) -- (.6,.6) ;
		\draw[cyan] (.3,-.7) -- (-.7,.3) ;
		\draw[cyan] (.7,-.3) -- (-.3,.7) ;
		\draw[black,fill=gray!15] (-.2,-.2) circle (.1 cm) ;
		\draw[black,fill=gray!15] (.2,.2) circle (.1 cm) ;
		\node at (.5,-.9) {\scriptsize 1} ;
		\node at (.9,-.5) {\scriptsize 2} ;
		\node at (-.9,.5) {\scriptsize 2} ;
		\node at (-.5,.9) {\scriptsize 1} ;
 \end{tikzpicture} =-\frac{1}{2!}\left( \prod_{i=1}^3 \int \frac{\dd E_i n_i}{4\pi E_i} \right) 32\pi i\,\ell_s^4\,E_1^2E_2^2E_3^3\,\delta(E_1-E_2)= i\ell_s^4\left(\frac{\zeta(3)^2}{\pi^2}-\frac{\zeta(3)}{6}\right) ,
\end{align}
with an identical contribution coming from the $l$-$r$ symmetric amplitude, giving a factor 2. The $2!$ comes from particles 1 and 2 being identical. Moving to the disconnected $T_{4{\to}4}$ contribution, whose expression is obtained from \reef{Mnm}, we find after collecting all numerical factors
\begin{align}\label{melon44}
\begin{tikzpicture}[line width=1.1 pt, scale=.7, baseline=(current bounding box.center)]
 		\draw[cyan] (-.5,-.5) -- (.5,.5) ;
		\draw[purple] (.5,-.5) -- (-.5,.5) ;
		\draw[cyan] (1,-.5) -- (2,.5) ;
		\draw[purple] (2,-.5) -- (1,.5) ;
		\draw[black,fill=gray!15] (0,0) circle (.1 cm) ;
		\draw[black,fill=gray!15] (1.5,0) circle (.1 cm) ;
		\node at (-.5,-.8) {\scriptsize 1} ;
		\node at (.5,-.8) {\scriptsize 2} ;
		\node at (1,-.8) {\scriptsize 3} ;
		\node at (2,-.8) {\scriptsize 4} ;
		\node at (-.5,.8) {\scriptsize 4} ;
		\node at (.5,.8) {\scriptsize 3} ;
		\node at (1,.8) {\scriptsize 2} ;
		\node at (2,.8) {\scriptsize 1} ;
 \end{tikzpicture} =-\frac{i\ell_s^4}{2\pi^2}\left(\frac{1}{2!}\right)^2\left( \prod_{i=1}^4 \int \dd E_i n_i E_i \right)\delta(E_1-E_3)\delta(E_2-E_4)= -\frac{i\ell_s^4}{2}\left(\frac{\pi}{6}-\frac{\zeta(3)}{\pi}\right)^2 .
\end{align}
This result has to be multiplied by two in order to take into account the history corresponding to the other non-vanishing clustering of $\{ 1,2,3,4 \}$, where 2 and 4 are exchanged in \reef{melon44}. For the remaining $T_{2{\to}2}T_{2{\to}2}$ diagram we need to multiply by 1/2 as dictated by the expansion of $\ln S(E)$, and get after collecting all numerical factors
\begin{align}\label{melon222}
\begin{tikzpicture}[line width=1.1 pt, scale=.7, baseline=(current bounding box.center)]
		\draw[cyan] (-.5,-.5) -- (.5,.5) ;
		\draw[purple] (.5,-.5) -- (-.5,.5) ;
		\draw[dashed,line width=.5] (-.85,.6) -- (.85,.6) ;
		\draw[cyan] (.5,.7) -- (-.5,1.7) ;
		\draw[purple] (-.5,.7) -- (.5,1.7) ;
		\draw[black,fill=gray!15] (0,0) circle (.1 cm) ;
		\draw[black,fill=gray!15] (0,1.2) circle (.1 cm) ;
 \end{tikzpicture} =\frac{i\ell_s^4}{2\pi^2}\,\frac{1}{2}\left( \prod_{i=1}^4 \int \dd E_i n_i E_i \right)e^{E_3+E_4}\,\delta(E_1-E_3)\delta(E_2-E_4)= \frac{i\ell_s^4\pi^2}{36}\, ,
\end{align}
where $e^{E_3+E_4}$ is there because the winding effects of internal particles 3 and 4 do not sum up to complete Bose-Einstein densities.
By inspecting the structure of \reef{melon33} to \reef{melon222}, we see no chance of getting something compatible with \reef{Esqrt}, primarily because the result is purely imaginary. When we combine everything we get a remarkable simplification
\be
2\times\reef{melon33}+2\times\reef{melon44}+\reef{melon222}=\frac{i\ell_s^4\zeta(3)^2}{\pi^2}\,.
\ee
Given that the free energy is a real object, we should expect this to be canceled by some other contribution. As a matter of fact, it turns out to be equal to the negative of the $O(\ell_s^4)$ contribution in \reef{flux2to2}, represented in \reef{vac24}. We can therefore conclude that \emph{the regular topologies do not contribute to the free energy of the Flux Tube at $O(\ell_s^4)$}.

Before moving to the singular topology, we would like to comment on the cancellation of the regular one. As it turns out, this is an example of a general property of the trace formula. Schematically, one finds that, at NLO, all the physics is contained in the real parts of the 1-loop $2{\to}2$ amplitude and tree-level $3{\to}3$, thanks to a delicate cancellation of the imaginary parts. These come from the unitarity cuts of the above mentioned amplitudes and from the disconnected single $T$ insertion and $T^2$ insertion. Interestingly enough, all the different densities nicely combine when taking the imaginary contributions, simplifying away and leaving
\be\label{regNLO}
f_{\rm NLO}=-\frac{1}{2!}\left( \prod_{i=1}^2\int \dd^d_{\rm th} k_i\right) \re M^{(1)}(1,2)-\frac{1}{3!}\left( \prod_{i=1}^3\int \dd^d_{\rm th} k_i\right) \re M^{(0)}(1,2,3)\,.
\ee
Armed with this, it is immediate to see now that the melon topology was bound to leave no contribution, because neither $M^{(1)}(1,2)$ nor the regular part of $M^{(0)}(1,2,3)$ have a real part.


\subsection{NLO correction: singular contribution}\label{sec:ls4sing}

Given that the non-singular topology gives zero when the various contributions are summed together, we expect the singular term of \reef{123to456}  to contain all the physical information of the $O(\ell_s^4)$ free energy \reef{Esqrt}. The point now is how to extract it, given that the forward limit is ill-defined. The key to resolving the singularity lies in the energy dependence of the operator $S(E)$ of \reef{master}. While for amplitudes that are regular in the forward limit it is enough to know that $\lim_{E\to E_\alpha}S_{\alpha\alpha}(E)=S_{\alpha\alpha}$, for singular amplitudes we need to know $S(E)$ in a neighborhood of $E_\alpha$.
When $T_{\alpha\alpha}(E)$ is convolved with a $\delta(E-E_\alpha)$, like in \reef{FofT}, it is tempting to do the  $\dd E$ integral and recover the physical forward amplitude. However, for forward-divergent  contributions to $T_{\alpha\alpha}$ this temptation should be tamed.

Qualitatively, the problem comes from diagrams with propagators that go on-shell in the forward limit. 
To   
include them in the trace formula, we need to  promote these diagrams $M_{\alpha\to\alpha}$ to their $E$-dependent counterpart $M_{\alpha\to\alpha}(E)$, as dictated by \reef{T_OFPT}. In order to do this, we have to refresh our OFPT toolkit, and in particular how to organize amplitudes with the Lippmann-Schwinger equation \reef{T_OFPT}. We show how this works focusing on the Flux Tube example, while trying to maintain generality as much as possible.

Starting from the singular contribution to \reef{123to456}, we proceed in steps: {\bf(1)} we rewrite the contribution that is singular in the forward limit as a sum of Feynman diagrams with covariant propagators; {\bf(2)} we decompose the so obtained expression the Old-Fashioned way, with $(E_\alpha-E_\gamma+i\epsilon)$ denominators; {\bf(3)} finally we promote the denominators according to $E_\alpha\to E$ and take the forward limit. As of {\bf (1)}, we have diagrammatically
\be\label{3to3Feynman}
\begin{tikzpicture}[line width=1.1 pt, scale=.7, baseline=(current bounding box.center)]
 		\draw[purple] (-.6,-.6) -- (.6,.6) ;
		\draw[cyan] (.3,-.7) -- (-.7,.3) ;
		\draw[cyan] (.7,-.3) -- (-.3,.7) ;
		\draw[black,fill=gray!15] (-.2,-.2) circle (.1 cm) ;
		\draw[black,fill=gray!15] (.2,.2) circle (.1 cm) ;
		\node at (.5,-.9) {\scriptsize 1} ;
		\node at (.9,-.5) {\scriptsize 2} ;
		\node at (-.9,.5) {\scriptsize 4} ;
		\node at (-.5,.9) {\scriptsize 5} ;
		\node at (-.9,-.7) {\scriptsize 3} ;
		\node at (.9,.7) {\scriptsize 6} ;
		
		\begin{scope}[shift={(4,0)}]
		\draw (-.6,-.6) -- (.6,.6) ;
		\draw (.3,-.7) -- (-.7,.3) ;
		\draw (.7,-.3) -- (-.3,.7) ;
		\node at (.5,-.9) {\scriptsize 1} ;
		\node at (.9,-.5) {\scriptsize 2} ;
		\node at (-.9,.5) {\scriptsize 4} ;
		\node at (-.5,.9) {\scriptsize 5} ;
		\node at (-.9,-.7) {\scriptsize 3} ;
		\node at (.9,.7) {\scriptsize 6} ;
		\end{scope}
		
		\begin{scope}[shift={(8,0)}]
		\draw (-.6,-.6) -- (.6,.6) ;
		\draw (.3,-.7) -- (-.7,.3) ;
		\draw (.7,-.3) -- (-.3,.7) ;
		\node at (.5,-.9) {\scriptsize 2} ;
		\node at (.9,-.5) {\scriptsize 1} ;
		\node at (-.9,.5) {\scriptsize 5} ;
		\node at (-.5,.9) {\scriptsize 4} ;
		\node at (-.9,-.7) {\scriptsize 3} ;
		\node at (.9,.7) {\scriptsize 6} ;
		\end{scope}
		
		\node at (2,0) {=} ;
		\node at (6,0) {+} ;		
 \end{tikzpicture}
\ee
where the graphs on the RHS are usual Feynman diagrams coming from the Nambu-Goto action.
Calling $Q=p_1+p_3-p_4$, we have in formulas
\be\label{3to3sing}
M_{1_l2_l3_r\to 4_l5_l6_r}^{(\rm sing)}
=-\,\ell_s^4\,\big( \,p_3{\cdot}p_1\ p_4^\mu+p_3{\cdot}p_4\ p_1^\mu\, \big)\,\frac{\,Q_\mu Q_\nu}{Q^2{+}i\epsilon}\,\big(\,p_6{\cdot}p_2\ p_5^\nu+p_6{\cdot}p_5\ p_2^\nu\,\big)+\big(1{\leftrightarrow}2,4{\leftrightarrow}5\big)\,.
\ee
In order to recover the first term of \reef{123to456}, it is useful to recall that $\frac{1}{x+i\epsilon}=\frac{\rm PV}{x}-i\pi\delta(x)$, and that in 1+1 dimensions $p_l{\cdot}p_r=2E_lE_r$, while $p_l{\cdot}q_l=p_r{\cdot}q_r=0$. Since the two Feynman diagrams of \reef{3to3Feynman} differ just by a relabelling of external legs, they contribute equally to the trace formula, so we focus on the first one and will multiply by two in due time. For step {\bf(2)} we employ the following identity among distributions, valid for any number of spacetime dimensions
\be\label{covtoOF}
\frac{\,Q_\mu Q_\nu}{Q^2{+}i\epsilon}=\delta_\mu^0\delta_\nu^0+\frac{1}{2|\vec{Q}|}\left(\,\frac{q_\mu\,q_\nu}{Q_0{-}|\vec{Q}|{+}i\epsilon}+\frac{\bar{q}_\mu\,\bar{q}_\nu}{{-}Q_0{-}|\vec{Q}|{+}i\epsilon}\,\right)\,,
\ee
where $q_\mu=(|\vec{Q}|,\vec{Q})$ and $\bar{q}_\mu=(|\vec{Q}|,-\vec{Q})$.~\footnote{Eq.~\reef{covtoOF} can be thought of as the unique decomposition of the analytic function in $Q_0$ on the LHS in terms of a polynomial $I(Q_0)$ that matches its behavior at infinity plus a sum over simple poles on the RHS.} Notice that the $\epsilon$ in the above formulas just means an infinitesimally small positive quantity, and one should not be worried by the fact that it has different dimensionality in the two sides of the equation. 

Interestingly, we see that on top of the usual time-ordered and anti-time-ordered contributions of OFPT amplitudes there is also a non-covariant contact term. The presence of this term is due to the $Q_\mu Q_\nu$ numerator, which comes from the derivative interactions of the Nambu-Goto action.

For the next step, we first need to rewrite the denominators of \reef{covtoOF} as the  difference of asymptotic and intermediate states energies
\begin{align}
\frac{1}{Q_0{-}|\vec{Q}|{+}i\epsilon}\ &=\ 
\begin{tikzpicture}[line width=1.1 pt, scale=.7, baseline=(current bounding box.center)]
 		\draw (-.8,-.8) -- (.8,.8) ;
		\draw (.2,-.8) -- (-.8,.2) ;
		\draw (.8,-.2) -- (-.2,.8) ;
		\draw[white,line width=2] (-1,0) -- (1,0) ;
		\draw[dashed,line width=.5] (-1,0) -- (1,0) ;
		\node at (.4,-1) {\scriptsize 1} ;
		\node at (1,-.4) {\scriptsize 2} ;
		\node at (-1,.4) {\scriptsize 4} ;
		\node at (-.4,1) {\scriptsize 5} ;
		\node at (-1,-1) {\scriptsize 3} ;
		\node at (1,1) {\scriptsize 6} ;
 \end{tikzpicture} \ =\ \frac{1}{(E_1{+}E_2{+}E_3)-(E_2{+}E_4{+}|\vec{Q}|)+i\epsilon} \label{OFtime} \\
 \frac{1}{-Q_0{-}|\vec{Q}|{+}i\epsilon}\ &=\ 
\begin{tikzpicture}[line width=1.1 pt, scale=.7, baseline=(current bounding box.center)]
 		\draw (-.8,-.8) -- (-.55,.4) -- (.55,-.4) -- (.8,.8) ;
		\draw (.2,-.8) -- (-.8,.8) ;
		\draw (.8,-.8) -- (-.2,.8) ;
		\draw[white,line width=2] (-1,0) -- (1,0) ;
		\draw[dashed,line width=.5] (-1,0) -- (1,0) ;
		\node at (.4,-1) {\scriptsize 1} ;
		\node at (1,-1) {\scriptsize 2} ;
		\node at (-1,1) {\scriptsize 4} ;
		\node at (-.4,1) {\scriptsize 5} ;
		\node at (-1,-1) {\scriptsize 3} ;
		\node at (1,1) {\scriptsize 6} ;
 \end{tikzpicture} \ =\ \frac{1}{(E_1{+}E_2{+}E_3)-(E_1{+}E_3{+}E_5{+}E_6{+}|\vec{Q}|)+i\epsilon} \label{OFanti}
\end{align}
where we used $Q_0=E_1{+}E_3{-}E_4=E_5{+}E_6{-}E_2$. Diagrams in \reef{OFtime} and \reef{OFanti} are OFPT diagrams: time flows upwards, and the dashed line denotes an intermediate state with all particles on the mass shell. Now step {\bf (3)} simply amounts to promoting $E_1{+}E_2{+}E_3{\to}E$ in \reef{OFtime} and \reef{OFanti} and plug back everything into \reef{3to3sing}.

Crucially, we see that the energy denominator in \reef{OFtime} is ill-defined in the forward limit, because $E_4=E_1$ and $|\vec{Q}|=E_3$. The anti-time-ordered denominator, on the contrary, has a smooth limit to $\frac{1}{-2E_3}$. Due to the ill behavior of the time-ordered diagram in the forward limit, we cannot simply take the amplitude $M_{\alpha\to\alpha}^{(\rm sing)}$ as in \reef{fofM}, and we should use $M_{\alpha\to\alpha}^{(\rm sing)}(E)$ instead. Given that $\bar{q}$ is a left-mover in the forward limit, the anti-time-ordered contribution vanishes, and we are left with
\begin{align}\label{123sing}
M_{1_l2_l3_r}^{(\rm sing)}(E)&=-8\,\ell_s^4\ p_1{\cdot}p_3 \ p_2{\cdot}p_3\ p_1^\mu \, p_2^\nu\left( \delta_\mu^0\delta_\nu^0+\frac{p_{3\mu}\, p_{3\nu}}{2E_3\big(E{-}( E_1{+}E_2{+}E_3){+}i\epsilon\big)}\, \right)\nonumber \\
&=-32\,\ell_s^4 \,E_1^2E_2^2E_3^2\left(1 +\frac{2E_3}{E{-}(E_1{+}E_2{+}E_3){+}i\epsilon}\right).
\end{align}
Because the $E$-dependent amplitude is still singular on the support of the $\delta(E{-}E_1{-}E_2{-}E_3)$, it might appear that we did not obtain much. However this is not true, as  can be  demonstrated by using  the following distributional identity
\be\label{delta'}
\delta(E-E_\alpha)\Re\frac{1}{E{-}E_\alpha{+}i\epsilon}=-\frac{1}{2}\partial_E\delta(E-E_\alpha)
\ee
which can be easily proven by using that $-2\pi i\delta(x){=}\frac{1}{x+i\epsilon}{-}\frac{1}{x-i\epsilon}$ and $\left(\frac{1}{x\pm i\epsilon}\right)^2{=}{-}\partial_x\frac{1}{x\pm i\epsilon}$. This identity  is  useful when --- in evaluating the trace formula --- we encounter a diagram with a forward-divergent OFPT denominator like \reef{123sing}, since we can now simply remove it according to
\be\label{trace_sing}
\int_0^\infty \dd E\,e^{-\beta E}\, \delta(E-E_\alpha)\,\Re\frac{N_\alpha}{E{-}E_\alpha{+}i\epsilon} = -\frac{\beta}{2} \,e^{-\beta E_\alpha}\,N_\alpha\, ,
\ee
where we did an integration by parts after using \reef{delta'}. $N_\alpha$ is a real numerator characterizing the $\alpha\to\alpha$ process, regular in the forward limit. The real part `Re' in \reef{trace_sing} is dictated by the trace formula, as was the case for Eq.~\reef{regNLO}.~\footnote{In the language of appendix~\ref{app_2level}, it arises from the interplay of the $O(T)$ term $VG_0V$ and the $O(T^2)$ term $-\frac{1}{2}V(G_0-\bar{G}_0)V$.}

It only remains to discuss windings. The two contributions in \reef{123sing} undergo a different resummation, in that the first term is enriched by windings of legs \{1,2,3\}, while the second has independent windings also for the intermediate state `$Q$'. This gives a factor of $e^{\beta E_Q}n_Q$ as explained in Sec.~\ref{sec:ls4reg}. In the forward limit $Q\to 3$, so we have
\begin{align}\label{f_llr}
f_{llr}=-\frac{1}{2!}\left(\prod_{i=1}^3\int \frac{\dd E_in_i}{4\pi E_i}\right)\left[\left(-32\ell_s^4E_1^2E_2^2E_3^2\right)-\frac{\beta}{2}\,e^{\beta E_3}n_3\left(-64\ell_s^4E_1^2E_2^2E_3^3\right)\right],
\end{align}
where the first and second term come respectively from the non-covariant contact term and the time-ordered contribution in \reef{123sing}. After evaluating the integral and adding the identical contribution from the $lrr$ configuration, we find
\be
f_{\rm NLO}=f_{llr}+f_{lrr}=-\frac{\ell_s^4\pi^3}{432\beta^6}\,,
\ee
in perfect agreement with \reef{Esqrt}. The key conceptual steps in establishing this result are basically two: on one side we needed to promote the forward-divergent contribution to a distribution in the extra $E$ variable, besides the kinematic ones; on the other side, the structure of the DMB equation \reef{master} is such that, when properly combining various singular contributions, all the spurious divergences go away, leaving a perfectly well-defined integral like \reef{f_llr}. We will see more examples of this mechanism calculating  higher orders in the next section.

\section{Higher orders}
\label{sec:higherorders}
\subsection{NNLO correction}\label{sec:ls6}

To compute NNLO effects we need to consider trace-connected diagrams with the following topologies
	\begin{equation}\label{vac46}
	\begin{tikzpicture}[line width=1.1 pt, scale=.7, baseline=(current bounding box.center)]
	
		\draw[cyan] (0,.8cm) -- (.692cm,-.4cm) -- (-.692cm,-.4cm) -- (0,.8cm) ;
		\draw[purple] (0,0) circle (.8cm) ;
		\draw[black,fill=gray!15] (0,.8cm) circle (.1 cm) ;
		\draw[black,fill=gray!15] (.692cm,-.4cm) circle (.1 cm) ;
		\draw[black,fill=gray!15] (-.692cm,-.4cm) circle (.1 cm) ;
		
		\begin{scope}[shift={(1.6,0)}]
		\draw[cyan] (4.8,0) circle (.6cm) ;
		\draw[cyan] (7.2,0) circle (.6cm) ;
		\draw[purple] (6,0) circle (.6cm) ;
		\draw[purple] (8.4,0) circle (.6cm) ;
		\draw[black,fill=gray!15] (5.4,0) circle (.1 cm) ;
		\draw[black,fill=gray!15] (6.6,0) circle (.1 cm) ;
		\draw[black,fill=gray!15] (7.8,0) circle (.1 cm) ;
		\end{scope}
		
		\begin{scope}[shift={(3.5,0)}]
		\draw[purple] (0,1.2) circle (.4cm) ;
		\draw[cyan] (0,0) circle (.8cm) ;
		\draw[purple] (0,0) ellipse (.8cm and .4cm) ;
		\draw[black,fill=gray!15] (0.77,0) circle (.1 cm) ;
		\draw[black,fill=gray!15] (-0.77,0) circle (.1 cm) ;
		\draw[black,fill=gray!15] (0,.8) circle (.1 cm) ;
		\end{scope}
		
		\begin{scope}[shift={(13,0)}]
		\draw[purple] (0,0) circle (.6cm) ;
		\draw[cyan] (0,1) circle (.4cm) ;
		\draw[cyan] (.866,-.5) circle (.4cm) ;
		\draw[cyan] (-.866,-.5) circle (.4cm) ;
		\draw[black,fill=gray!15] (0,.6cm) circle (.1 cm) ;
		\draw[black,fill=gray!15] (.519cm,-.3cm) circle (.1 cm) ;
		\draw[black,fill=gray!15] (-.519cm,-.3cm) circle (.1 cm) ;
		\end{scope}
		
	 \end{tikzpicture}
	 \end{equation}
together with the lower order diagrams \reef{vac34} and \reef{vac24} with higher order vertices ($\ell_s^4$ and $\ell_s^6$). At NLO we saw that only the singular topology contributed. It was a priori apparent that the other diagrams were bound to cancel with each other. Not just because they were purely imaginary, but also because the powers of $\pi$ from these diagrams   were not matching the expected result from the expansion of \reef{Esqrt}. We expect something analogous to happen here. As an example, consider the contribution coming from tracing the $O(\ell_s^6)$ term of \reef{flux2to2}, yielding
\be
\begin{tikzpicture}[line width=1.1 pt, scale=.8, baseline=(current bounding box.center)]
		\draw[purple] (-.5,-.5) -- (.5,.5) ;
		\draw[cyan] (.5,-.5) -- (-.5,.5) ;
		\draw[black,fill=gray!15] (0,0) circle (.1 cm) ;
		\node at (0.6,0) {\scriptsize $\ell_s^6$} ;
		\draw[white] (0,-.6) -- (0,-.7) ;
		\node at (-.58,-.75) {\tiny 1} ;
		\node at (-.55,.75) {\tiny 2} ;
		\node at (.55,-.75) {\tiny 2} ;
		\node at (.58,.75) {\tiny 1} ;
		\end{tikzpicture}  = \frac{\ell_s^6}{12\pi^2}\left(\int_0^\infty \dd E\,n(E) E^3\right)^2=\frac{\ell_s^6}{12\pi^2}\left(\frac{\pi^4}{15}\right)^2=\frac{\ell_s^6\pi^6}{2700}\,.
\ee
Despite being real (as $i^2$ is), it does not have the correct $\pi$ power, and we expect it to cancel with diagrams that have no singular propagators, like for the `melon' topology in the previous section. The non-singular topologies at this order are the first of \reef{vac46} and the melon of the previous section with an $O(\ell_s^4)$ vertex, together with the one we just computed. There is also a topology corresponding to amplitudes with a single singular propagator (in the forward limit), represented by the second graph of \reef{vac46}. When the singular propagator is regulated like in \reef{trace_sing}, these diagrams give a purely imaginary contribution and they are bound to cancel among themselves due to the reality of the trace formula \reef{master}.

We are going to indirectly show the cancellation of these topologies by calculating the contribution of the maximally-singular topologies, i.e. the last two of \reef{vac46}, and show that the result saturates the $O(\ell_s^6)$ of \reef{Esqrt}.

\subsubsection{$lllr$ \& $lrrr$ topologies}

We start by evaluating the last topology of \reef{vac46}, which in terms of amplitudes corresponds to the maximally singular contribution to the forward limit of $lllr\to lllr$, called from here on $M_{1_l2_l3_l4_r}^{\rm (sing)}$. There are now two singular propagators. In order to regulate them, like before, we need first of all to calculate $M_{1_l2_l3_l4_r}^{\rm (sing)}(E)$, the amplitude with shifted energy $E$. A crucial step to compute it, analogously to what we did before, is to establish the following identity
\be\label{3l3rperm}
\begin{tikzpicture}[line width=1.1 pt, scale=.7, baseline=(current bounding box.center)]
 		\draw[purple] (-.6,-.6) -- (1,1) ;
		\draw[cyan] (.3,-.7) -- (-.7,.3) ;
		\draw[cyan] (.7,-.3) -- (-.3,.7) ;
		\draw[cyan] (1.1,.1) -- (.1,1.1) ;
		\draw[black,fill=gray!15] (-.2,-.2) circle (.1 cm) ;
		\draw[black,fill=gray!15] (.2,.2) circle (.1 cm) ;
		\draw[black,fill=gray!15] (.6,.6) circle (.1 cm) ;
		\node at (.5,-.9) {\scriptsize 1} ;
		\node at (.9,-.5) {\scriptsize 2} ;
		\node at (1.3,-.1) {\scriptsize 3} ;
		\node at (-.9,.5) {\scriptsize 5} ;
		\node at (-.5,.9) {\scriptsize 6} ;
		\node at (-.1,1.3) {\scriptsize 7} ;
		\node at (-.9,-.7) {\scriptsize 4} ;
		\node at (1.3,1.1) {\scriptsize 8} ;
		
		\begin{scope}[shift={(5,0)}]
		\draw (-.6,-.6) -- (1,1) ;
		\draw (.3,-.7) -- (-.7,.3) ;
		\draw (.7,-.3) -- (-.3,.7) ;
		\draw (1.1,.1) -- (.1,1.1) ;
		\node at (.5,-.9) {\scriptsize 1} ;
		\node at (.9,-.5) {\scriptsize 2} ;
		\node at (1.3,-.1) {\scriptsize 3} ;
		\node at (-.9,.5) {\scriptsize 5} ;
		\node at (-.5,.9) {\scriptsize 6} ;
		\node at (-.1,1.3) {\scriptsize 7} ;
		\node at (-.9,-.7) {\scriptsize 4} ;
		\node at (1.3,1.1) {\scriptsize 8} ;
		\end{scope}
		
		\node at (2.6,0.2) {=} ;
		
		\node at (10,0.2) {+ ~~ 5 permutations} ;
	
 \end{tikzpicture}
\ee
where the permutations keep fixed 4 and 8, and are such that 1 and 5 always join the same vertex, and similarly for 2,6 and 3,7. Under the trace all 6 Feynman diagrams are equivalent, so we can focus on the first. Structurally, we have
\be\label{lllrFeyn}
\begin{tikzpicture}[line width=1.1 pt, scale=.7, baseline=(current bounding box.center)]

		\draw (-.6,-.6) -- (1,1) ;
		\draw (.3,-.7) -- (-.7,.3) ;
		\draw (.7,-.3) -- (-.3,.7) ;
		\draw (1.1,.1) -- (.1,1.1) ;
		\node at (.5,-.9) {\scriptsize 1} ;
		\node at (.9,-.5) {\scriptsize 2} ;
		\node at (1.3,-.1) {\scriptsize 3} ;
		\node at (-.9,.5) {\scriptsize 5} ;
		\node at (-.5,.9) {\scriptsize 6} ;
		\node at (-.1,1.3) {\scriptsize 7} ;
		\node at (-.9,-.7) {\scriptsize 4} ;
		\node at (1.3,1.1) {\scriptsize 8} ;
	
 \end{tikzpicture}  = \ A_\mu B_\nu C_{\rho\sigma}  \, \frac{\,P^\mu P^\rho}{P^2{+}i\epsilon} \ \frac{\,Q^\nu Q^\sigma}{Q^2{+}i\epsilon}
\ee
where $A,B$ and $C$ depend only on the external kinematics, $P=p_1{+}p_4{-}p_5$ and $Q=p_7{+}p_8{-}p_3$. 
Starting from \reef{lllrFeyn}, we can use \reef{covtoOF} twice to rewrite the covariant propagators in terms of energy denominators and non-covariant contact terms. Similarly to the NLO case, in the forward limit only the time-ordered and contact pieces remain. Pieces having at least one anti-time-ordered  denominator vanish when contracted to the external kinematics, so we can write
\be
\begin{tikzpicture}[line width=1.1 pt, scale=.7, baseline=(current bounding box.center)]

		\draw (-.6,-.6) -- (1,1) ;
		\draw (.3,-.7) -- (-.7,.3) ;
		\draw (.7,-.3) -- (-.3,.7) ;
		\draw (1.1,.1) -- (.1,1.1) ;
		\node at (.5,-.9) {\scriptsize 1} ;
		\node at (.9,-.5) {\scriptsize 2} ;
		\node at (1.3,-.1) {\scriptsize 3} ;
		\node at (-.9,.5) {\scriptsize 5} ;
		\node at (-.5,.9) {\scriptsize 6} ;
		\node at (-.1,1.3) {\scriptsize 7} ;
		\node at (-.9,-.7) {\scriptsize 4} ;
		\node at (1.3,1.1) {\scriptsize 8} ;
	
 \end{tikzpicture}  = \ A^\mu B^\nu C^{\rho\sigma}  \left(\,\delta_\mu^0\delta_\rho^0 + \frac{p_\mu\,p_\rho}{P_0{-}|\vec{P}|{+}i\epsilon}\, \right)\left(\,\delta_\nu^0\delta_\sigma^0 + \frac{q_\nu\,q_\sigma}{Q_0{-}|\vec{Q}|{+}i\epsilon}\,\right)+\ldots
\ee
where the ellipses stand for terms that vanish in the forward limit. The treatment of terms with either one or no energy denominator proceeds exactly as before. On the other hand, the term with two energy denominators is new, so let us comment on it. As emphasized previously, the important question is whether the denominators can be written as \emph{energy differences} among asymptotic and intermediate states, or equivalently if they correspond to an OFPT diagram with two intermediate states, something we should not take for granted, as we explain in the next section. The answer in this case is positive, and we find (omitting the numerators)
\be\label{nnlo1OFPT}
\frac{1}{P_0{-}|\vec{P}|{+}i\epsilon}\,\frac{1}{Q_0{-}|\vec{Q}|{+}i\epsilon}~ \equiv~
\begin{tikzpicture}[line width=1.1 pt, scale=.7, baseline=(current bounding box.center)]

		\draw (-.6,-.6) -- (1,1) ;
		\draw (.3,-.7) -- (-.7,.3) ;
		\draw (.7,-.3) -- (-.3,.7) ;
		\draw (1.1,.1) -- (.1,1.1) ;
		\draw[white,line width=2] (-1,0) -- (1,0) ;
		\draw[dashed,line width=.5] (-1,0) -- (1,0) ;
		\draw[white,line width=2] (-.6,0.4) -- (1.4,0.4) ;
		\draw[dashed,line width=.5] (-.6,0.4) -- (1.4,0.4) ;
		\node at (.5,-.9) {\scriptsize 1} ;
		\node at (.9,-.5) {\scriptsize 2} ;
		\node at (1.3,-.1) {\scriptsize 3} ;
		\node at (-.9,.5) {\scriptsize 5} ;
		\node at (-.5,.9) {\scriptsize 6} ;
		\node at (-.1,1.3) {\scriptsize 7} ;
		\node at (-.9,-.7) {\scriptsize 4} ;
		\node at (1.3,1.1) {\scriptsize 8} ;
	
 \end{tikzpicture}~=~\frac{1}{(E_\alpha-E_\gamma+i\epsilon)(E_\alpha-E_{\gamma'}+i\epsilon)}
\ee
where, in the notation of \reef{T_OFPT}, we have $E_\alpha=E_1{+}E_2{+}E_3{+}E_4$, while the intermediate states energies are $E_\gamma=E_2{+}E_3{+}E_5{+}|\vec{P}|$ and $E_{\gamma'}=E_3{+}E_5{+}E_6{+}|\vec{Q}|$. Notice that, like for \reef{OFtime}, the topology of \reef{nnlo1OFPT} is such that $E_\gamma,E_{\gamma'}\to E_\alpha$ in the forward limit, so the diagram is singular. Before taking the limit, we have then to promote $E_\alpha\to E$ as dictated by \reef{master}.

Using that, in the forward limit,
\begin{align}
A^\mu=4\ell_s^2 \,E_1E_4 \,p_1^\mu 
~~~,~~~~~ B^\nu=4\ell_s^2 \,E_3E_4 \,p_3^\nu
~~~,~~~~~C^{\rho\sigma}=2\ell_s^2 \, p_2^\rho\, p_2^\sigma
\end{align}
and that $P,Q,p,q\to p_4$, we find after a little algebra
\be\label{4to4sing}
M_{1_l2_l3_l4_r}^{\rm (sing)}(E)=3!\,32\ell_s^6\, \prod_{i=1}^4E_i^2\left(1+\frac{4E_4}{E{-}E_{\alpha}{+}i\epsilon}+ \frac{4E_4^2}{\left(E{-}E_{\alpha}{+}i\epsilon\right)^2}\right).
\ee
Like for the NLO, we have now to plug \reef{4to4sing} into the trace formula, taking care of adding {\bf (i)} $O(T^2)$ and $O(T^3)$ terms and {\bf (ii)} windings. Regarding {\bf (i)}, we are now going to show how the discussion below \reef{trace_sing} generalizes to higher orders.

The forward amplitude \reef{4to4sing} is of the form $M(E)=M^{(1)}+g_0M^{(2)}+g_0^2M^{(3)}$, where $g_0\equiv(E{-}E_{\alpha}{+}i\epsilon)^{-1}$, and $M^{(i)}$ are independent of $E$. Following \reef{fofM}, it contributes to the trace formula as 
\begin{align}\label{Fsingular}
\Delta f =-\int_0^\infty \dd E e^{-\beta E} \int \dd \alpha \,\delta(E-E_\alpha)\left[ M^{(1)}+g_0M^{(2)}+g_0^2M^{(3)}\right]+\ldots 
\end{align}
where the ellipses stand for terms with more $T$ insertions. In order to capture these contributions, we expand the logarithm of \reef{master} to higher orders in $T$ (for the case at hand, up to $O(T^3)$) and find after inspection that there exist contributions identical to \reef{Fsingular}, except for having Dirac deltas instead of energy denominators. Rewriting deltas for convenience according to $g_0{-}\bar{g}_0=-2\pi i\delta(E-E_\alpha)$, we find after adding the new contributions
\begin{align}\label{FsingT}
\Delta f= \int_0^\infty  \dd E e^{-\beta E} \int \dd \alpha \ \frac{g_0-\bar{g}_0}{2\pi i}\,\left[ M^{(1)}+\frac{1}{2}(g_0{+}\bar{g}_0)\,M^{(2)}+\frac{1}{3}\left(g_0^2{+}g_0\bar{g}_0{+}\bar{g}_0^2\right)M^{(3)}\right].
\end{align}
To prove this, we start from an expansion of $\ln S(E)$ in powers of $V$, with $T=V+VG_0V+\ldots$ thanks to the Lippmann-Schwinger equation \reef{T_OFPT} and, after a couple of simplifications, we find
\begin{align}\label{lnSV3}
\ln S(E)=(G_0{-}\bar{G}_0)\bigg[ V+\frac{1}{2}\Re VG_0V+\frac{2}{3}\Re VG_0V{G}_0V+\frac{1}{3}\Re VG_0V\bar{G}_0V\bigg]+O(V^4)\,.
\end{align}
In order to go from this to \reef{FsingT}, the key is to focus on contributions to \reef{master} where the intermediate states $\alpha$ that are inserted into \reef{lnSV3} to evaluate the operator products (via $\int |\alpha\rangle\langle \alpha|$) are exactly coincident with the `asymptotic' ones used to take the trace. Under this restriction, all resolvents, which become now c-numbers, are equal to $g_0$ and can be collected together, leading to \reef{FsingT}. As we showed explicitly at NLO, these contributions to the trace are precisely those that feature singular propagators in the forward limit and, for the integrable Nambu-Goto theory, they are actually the only ones that matter!

When we combine $g_0-\bar{g}_0$ with the other resolvents in \reef{FsingT}, they nicely simplify thanks to
\be
g_0^n-\bar{g}_0^n=2\pi i\frac{(-1)^{n}}{(n-1)!}\partial_E^{n-1}\delta(E-E_\alpha)\,.
\ee
Integrating over $\dd E$ using partial integration to let derivatives act on the Boltzmann factor, we get 
\be\label{OV3simple}
\Delta f = - \int \dd \alpha  \,e^{-\beta E_\alpha}\left[ M^{(1)}-\frac{\beta}{2}M^{(2)}+\frac{\beta^2}{6}M^{(3)}\right].
\ee
It is clear that this pattern continues, with terms of $O(\beta^n)$ as a leftover of would-be singular contributions having $n$ propagators going on-shell. We refrain from giving more details of the generalization here. 

We are now finally in a position to add windings. For the topology at hand, we find
\be\label{flllr}
 f_{lllr} = -\frac{1}{3!}\left(\prod_{i=1}^4\int \frac{\dd E_in_i}{4\pi E_i}\right) \left[ M^{(1)}-\frac{\beta}{2}n_4 e^{\beta E_4}M^{(2)}+\frac{\beta^2}{6}n_4^2(e^{2\beta E_4}+e^{\beta E_4})M^{(3)}\right]
\ee
which can be easily evaluated using \reef{4to4sing}, to obtain
\be
f_{lllr}=-\frac{\ell_s^6\pi^4}{10368\,\beta^8}(1-4+4)=-\frac{\ell_s^6\pi^4}{10368\,\beta^8}\,,
\ee
with an identical yield coming from $f_{lrrr}$. The factors in brackets come from the various contributions proportional to $M^{(i)}$.

The effect of windings for the term proportional to $M^{(3)}$ is surely not obvious to guess. However it is obtained, as before, by carefully summing over the independent windings of the four external legs and the two propagators $P$ and $Q$, identifying both propagators with $p_4$ in the forward limit at the end. We observe that the final expression can be rewritten as $n_1n_2n_3n''_4$, while the winding of the $M^{(2)}$ contribution is $n_1n_2n_3n'_4$. The rule of thumb is that we get as many derivatives on $n_i$ as the number of on-shell propagators that particle $i$ `traverses'.

\subsubsection{$llrr$ topology}

With this section we conclude the computation of NNLO effects to the integrable Nambu-Goto theory free energy. The logic is the same as before: we start from the factorized amplitude, perform an OFPT `scan' of it, and finally integrate it after removing the divergent energy denominators according to \reef{OV3simple}. There are two notable novelties for this topology
\begin{enumerate}
\item When trying to express the integrable amplitude in terms of Feynman diagrams with two propagators like in \reef{lllrFeyn}, we find it necessary to add a compensating covariant contact term.
\be\label{4to4Feyn}
\begin{tikzpicture}[line width=1.1 pt, scale=.33, baseline=(current bounding box.center)]

		\draw[purple] (-1,-2) -- (2,1) ;
		\draw[purple] (-2,-1) -- (1,2) ;
		\draw[cyan] (1,-2) -- (-2,1) ;
		\draw[white,line width=4] (2,-1) -- (-1,2) ;
		\draw[cyan] (2,-1) -- (-1,2) ;
		\draw[black,fill=gray!15] (0,-1) circle (.2 cm) ;
		\draw[black,fill=gray!15] (1,0) circle (.2 cm) ;
		\draw[black,fill=gray!15] (-1,0) circle (.2 cm) ;
		\node at (-1.3,-2.3) {\scriptsize 3} ;
		\node at (-2.3,-1.3) {\scriptsize 4} ;
		\node at (1.3,-2.3) {\scriptsize 1} ;
		\node at (2.3,-1.3) {\scriptsize 2} ;
		\node at (-1.3,2.3) {\scriptsize 6} ;
		\node at (-2.3,1.3) {\scriptsize 5} ;
		\node at (1.3,2.3) {\scriptsize 8} ;
		\node at (2.3,1.3) {\scriptsize 7} ;
		
		\node at (4,0) {=} ;
		
		\begin{scope}[shift={(8,0)}]
		\draw (-1,-2) -- (2,1) ;
		\draw (-2,-1) -- (1,2) ;
		\draw (1,-2) -- (-2,1) ;
		\draw[white,line width=4] (.5,.5) -- (-1,2) ;
		\draw (2,-1) -- (-1,2) ;
		\draw (0,-1) circle (.15);
		\draw (1,0) circle (.15);
		\draw (-1,0) circle (.15);
		\node at (-1.3,-2.3) {\scriptsize 3} ;
		\node at (-2.3,-1.3) {\scriptsize 4} ;
		\node at (1.3,-2.3) {\scriptsize 1} ;
		\node at (2.3,-1.3) {\scriptsize 2} ;
		\node at (-1.3,2.3) {\scriptsize 6} ;
		\node at (-2.3,1.3) {\scriptsize 5} ;
		\node at (1.3,2.3) {\scriptsize 8} ;
		\node at (2.3,1.3) {\scriptsize 7} ;
		\end{scope}
		
		\begin{scope}[shift={(15,0)}]
		\draw (-1,-2) -- (2,1) ;  
		\draw (-2,-1) -- (1,2) ; 
		\draw[white,line width=4] (-.5,-.5) -- (-2,1) ;
		\draw (1,-2) -- (-2,1) ; 
		\draw (2,-1) -- (-1,2) ; 
		\draw (0,-1) circle (.15);
		\draw (1,0) circle (.15);
		\draw (0,1) circle (.15);
		\node at (-1.3,-2.3) {\scriptsize 3} ;
		\node at (-2.3,-1.3) {\scriptsize 4} ;
		\node at (1.3,-2.3) {\scriptsize 2} ;
		\node at (2.3,-1.3) {\scriptsize 1} ;
		\node at (-1.3,2.3) {\scriptsize 5} ;
		\node at (-2.3,1.3) {\scriptsize 6} ;
		\node at (1.3,2.3) {\scriptsize 8} ;
		\node at (2.3,1.3) {\scriptsize 7} ;
		\end{scope}
		
		\begin{scope}[shift={(22,0)}]
		\draw (2,-1) -- (-1,2) ; 
		\draw[white,line width=4] (.5,-.5) -- (2,1) ;
		\draw (-1,-2) -- (2,1) ;  
		\draw (-2,-1) -- (1,2) ; 
		\draw (1,-2) -- (-2,1) ; 
		\draw (0,-1) circle (.15);
		\draw (-1,0) circle (.15);
		\draw (0,1) circle (.15);
		\node at (-1.3,-2.3) {\scriptsize 4} ;
		\node at (-2.3,-1.3) {\scriptsize 3} ;
		\node at (1.3,-2.3) {\scriptsize 1} ;
		\node at (2.3,-1.3) {\scriptsize 2} ;
		\node at (-1.3,2.3) {\scriptsize 6} ;
		\node at (-2.3,1.3) {\scriptsize 5} ;
		\node at (1.3,2.3) {\scriptsize 7} ;
		\node at (2.3,1.3) {\scriptsize 8} ;
		\end{scope}
		
		\begin{scope}[shift={(29,0)}]
		\draw (-2,-1) -- (1,2) ;
		\draw (1,-2) -- (-2,1) ;
		\draw[white,line width=4] (.5,-.5) -- (-1,-2) ;
		\draw (2,-1) -- (-1,2) ;
		\draw (-1,-2) -- (2,1) ;
		\draw (-1,0) circle (.15);
		\draw (1,0) circle (.15);
		\draw (0,1) circle (.15);
		\node at (-1.3,-2.3) {\scriptsize 4} ;
		\node at (-2.3,-1.3) {\scriptsize 3} ;
		\node at (1.3,-2.3) {\scriptsize 2} ;
		\node at (2.3,-1.3) {\scriptsize 1} ;
		\node at (-1.3,2.3) {\scriptsize 5} ;
		\node at (-2.3,1.3) {\scriptsize 6} ;
		\node at (1.3,2.3) {\scriptsize 7} ;
		\node at (2.3,1.3) {\scriptsize 8} ;
		\end{scope}
		
		\begin{scope}[shift={(1.5,0)}]
		\node at (10,0) {+};
		\node at (17,0) {+};
		\node at (24,0) {+};
		\end{scope}
		
		\draw[white] (0,-2) -- (0,-3.5) ;

 \end{tikzpicture} -~32\ell_s^6 \prod_{i=1}^8E_i\,.
\ee
The first diagram on the RHS is given by
\be\label{2l2rFeyn}
\begin{tikzpicture}[line width=1.1 pt, scale=.33, baseline=(current bounding box.center)]

		\draw (-1,-2) -- (2,1) ;
		\draw (-2,-1) -- (1,2) ;
		\draw (1,-2) -- (-2,1) ;
		\draw[white,line width=4] (.5,.5) -- (-1,2) ;
		\draw (2,-1) -- (-1,2) ;
		
		\draw (0,-1) circle (.15);
		\draw (1,0) circle (.15);
		\draw (-1,0) circle (.15);
		
		\node at (-1.3,-2.3) {\tiny 3} ;
		\node at (-2.3,-1.3) {\tiny 4} ;
		\node at (1.3,-2.3) {\tiny 1} ;
		\node at (2.3,-1.3) {\tiny 2} ;
		\node at (-1.3,2.3) {\tiny 6} ;
		\node at (-2.3,1.3) {\tiny 5} ;
		\node at (1.3,2.3) {\tiny 8} ;
		\node at (2.3,1.3) {\tiny 7} ;
		
		\draw[white] (0,-2) -- (0,-3.3) ;
		
\end{tikzpicture}
~=~32\ell_s^6  \frac{E_5 E_7 }{(E_8{-}E_4{+}i\epsilon)(E_6{-}E_2{+}i\epsilon)}\prod_{i=1}^8E_i\,,
\ee
and the other diagrams have similar expressions: instead of $E_5E_7$ in the numerator, we find $E_3E_5$, $E_1E_7$ and $E_1E_3$ for respectively the second, third and fourth diagram. The denominator structure is identical up to signs. To show the validity of \reef{4to4Feyn}, it is useful to decompose the propagators into a principal value and a Dirac delta. Terms containing a single delta cancel among themselves, while those with no delta sum up to a constant term equal to $+32\ell_s^6 \prod_{i=1}^8E_i$.
\item There is a new subtlety in doing the OFPT analysis for two out of the four covariant diagrams of \reef{4to4Feyn}, specifically the first and the last. As we are going to show, the repeated use of \reef{covtoOF} does not give directly a proper OFPT decomposition of these two diagrams, and we need further massaging.
\end{enumerate}
We start from a rewriting of \reef{2l2rFeyn} of the form
\be\label{llrrFeyn}
\begin{tikzpicture}[line width=1.1 pt, scale=.33, baseline=(current bounding box.center)]

		\draw (-1,-2) -- (2,1) ;
		\draw (-2,-1) -- (1,2) ;
		\draw (1,-2) -- (-2,1) ;
		\draw[white,line width=4] (.5,.5) -- (-1,2) ;
		\draw (2,-1) -- (-1,2) ;
		
		\draw (0,-1) circle (.15);
		\draw (1,0) circle (.15);
		\draw (-1,0) circle (.15);
		
		\node at (-1.3,-2.3) {\tiny 3} ;
		\node at (-2.3,-1.3) {\tiny 4} ;
		\node at (1.3,-2.3) {\tiny 1} ;
		\node at (2.3,-1.3) {\tiny 2} ;
		\node at (-1.3,2.3) {\tiny 6} ;
		\node at (-2.3,1.3) {\tiny 5} ;
		\node at (1.3,2.3) {\tiny 8} ;
		\node at (2.3,1.3) {\tiny 7} ;
		
		\draw[white] (0,-2) -- (0,-3.3) ;
	
 \end{tikzpicture}  = \ A_\mu B_\nu C_{\rho\sigma}  \, \frac{\,P^\mu P^\rho}{P^2{+}i\epsilon} \ \frac{\,Q^\nu Q^\sigma}{Q^2{+}i\epsilon}
\ee
where $P=p_5{+}p_8{-}p_4$, $Q=p_6{+}p_7{-}p_2$ and $A,B,C$ have smooth forward limit, specifically $A^\mu\to 4\ell_s^2 E_1E_4 p_4^\mu$, $B^\nu\to 4\ell_s^2 E_2E_3 p_2^\nu$ and $C^{\rho\sigma}\to 2\ell_s^2 p_3^\rho p_4^\sigma$. Like for the $lllr$ topology, due to the numerator structure only the time-ordered and contact terms survive the forward limit. However, unlike before, the term with a product of two energy denominators cannot be interpreted right away as an OFPT diagram.

To understand what is going on, we recall that for each covariant diagram with $n$ vertices there are $n!$ OFPT diagrams, obtained by adding all possible time orderings of the vertices \cite{Weinberg:1966jm}. Two covariant propagators require 6 OFPT orderings to be reproduced, but the repeated application of \reef{covtoOF} only has 4 terms (we are tacitly excluding the contact terms from this discussion). Let us focus for concreteness on the term that survives the forward limit of \reef{llrrFeyn}, whose denominator is $(P_0{-}|\vec{P}|{+}i\epsilon)(Q_0{-}|\vec{Q}|{+}i\epsilon)$. The time ordering implicit in each of the two factors implies that {\bf (1)} $t_{1\,{\rm meets}\,3}<t_{2\,{\rm meets}\,3}$ and {\bf (2)} $t_{1\,{\rm meets}\,3}<t_{1\,{\rm meets}\,4}$, but the ordering among $t_{2\,{\rm meets}\,3}$ and $t_{1\,{\rm meets}\,4}$ is not resolved. For comparison, the LHS of \reef{nnlo1OFPT} corresponds instead to $t_{1\,{\rm meets}\,4}<t_{2\,{\rm meets}\,4}<t_{3\,{\rm meets}\,4}$, all 3 events being ordered.

The two possible orderings among $t_{2\,{\rm meets}\,3}$ and $t_{1\,{\rm meets}\,4}$ correspond, first in diagrams and then in formulas, to
\be\label{2orderings}
\begin{tikzpicture}[line width=1 pt, scale=.23, baseline=(current bounding box.center)]

	\begin{scope}[shift={(9,0)}]
	\draw (-3,4) -- (1,0) ;
	\draw (1,0) -- (3,2) ;
	\draw[dashed, line width = .5] (-3.5,3) -- (1,3) ;
	\draw[dashed, line width = .5] (-2,1) -- (4,1) ;
	\node at (1,0) {\scriptsize$\bullet$} ;
	\node at (-3,4) {\scriptsize$\bullet$} ;
	\node at (3,2) {\scriptsize$\bullet$} ;
	\end{scope}
	
	\node at (4.5,1.5) {+} ;
	
	\begin{scope}[shift={(0,0)}]
	\draw (-3,2) -- (-1,0) ;
	\draw (-1,0) -- (3,4) ;
	\draw[dashed, line width = .5] (3.5,3) -- (-1,3) ;
	\draw[dashed, line width = .5] (2,1) -- (-4,1) ;
	\node at (-1,0) {\scriptsize$\bullet$} ;
	\node at (-3,2) {\scriptsize$\bullet$} ;
	\node at (3,4) {\scriptsize$\bullet$} ;
	\end{scope}
\end{tikzpicture}
~~=\frac{1}{Q_0{+}P_0{-}|{\vec Q}|{-}|{\vec P}|{+}i\epsilon}\left(\frac{1}{P_0{-}|{\vec P}|{+}i\epsilon}+\frac{1}{Q_0{-}|{\vec Q}|{+}i\epsilon}\right) \nonumber
\ee
\be
=\frac{1}{(P_0{-}|\vec{P}|{+}i\epsilon)(Q_0{-}|\vec{Q}|{+}i\epsilon)}~\,
\ee

\noindent the two denominator products in the line above being dictated by which legs are cut by the dashed lines. When we now embed \reef{2orderings} into \reef{llrrFeyn}, we find that each of the two denominator products is of the correct form $(E_\alpha-E_\gamma+i\epsilon)(E_\alpha-E_{\gamma'}+i\epsilon)$, and, after we promote $E_\alpha\to E$, \emph{they have the same forward limit} $(E-E_{\alpha}+i\epsilon)^2$.

Putting together all the ingredients, we finally find
\be\label{2L2Rforward1}
\lim_{{\rm forward}'}
\begin{tikzpicture}[line width=1.1 pt, scale=.33, baseline=(current bounding box.center)]

		\draw (-1,-2) -- (2,1) ;
		\draw (-2,-1) -- (1,2) ;
		\draw (1,-2) -- (-2,1) ;
		\draw[white,line width=4] (.5,.5) -- (-1,2) ;
		\draw (2,-1) -- (-1,2) ;
		
		\draw (0,-1) circle (.15);
		\draw (1,0) circle (.15);
		\draw (-1,0) circle (.15);
		
		\node at (-1.3,-2.3) {\tiny 3} ;
		\node at (-2.3,-1.3) {\tiny 4} ;
		\node at (1.3,-2.3) {\tiny 1} ;
		\node at (2.3,-1.3) {\tiny 2} ;
		\node at (-1.3,2.3) {\tiny 6} ;
		\node at (-2.3,1.3) {\tiny 5} ;
		\node at (1.3,2.3) {\tiny 8} ;
		\node at (2.3,1.3) {\tiny 7} ;
		
		\draw[white] (0,-2) -- (0,-3.3) ;
	
 \end{tikzpicture}  = 32\ell_s^6\, \prod_{i=1}^4E_i^2\left(1+\frac{2E_1+2E_3}{E{-}E_{\alpha}{+}i\epsilon}+ \frac{8E_1E_3}{\left(E{-}E_{\alpha}{+}i\epsilon\right)^2}\right),
\ee
where forward$'$ means that we shift to $E$ before taking the forward limit. For contrast, the second diagram of \reef{4to4Feyn} gives
\be\label{2L2Rforward2}
\lim_{{\rm forward}'}
\begin{tikzpicture}[line width=1.1 pt, scale=.33, baseline=(current bounding box.center)]

		\draw (-1,-2) -- (2,1) ;  
		\draw (-2,-1) -- (1,2) ; 
		\draw[white,line width=4] (-.5,-.5) -- (-2,1) ;
		\draw (1,-2) -- (-2,1) ; 
		\draw (2,-1) -- (-1,2) ; 
		
		\draw (0,-1) circle (.15);
		\draw (1,0) circle (.15);
		\draw (0,1) circle (.15);
		
		\node at (-1.3,-2.3) {\tiny 3} ;
		\node at (-2.3,-1.3) {\tiny 4} ;
		\node at (1.3,-2.3) {\tiny 1} ;
		\node at (2.3,-1.3) {\tiny 2} ;
		\node at (-1.3,2.3) {\tiny 6} ;
		\node at (-2.3,1.3) {\tiny 5} ;
		\node at (1.3,2.3) {\tiny 8} ;
		\node at (2.3,1.3) {\tiny 7} ;
		
		\draw[white] (0,-2) -- (0,-3.3) ;
	
 \end{tikzpicture}  = 32\ell_s^6\, \prod_{i=1}^4E_i^2\left(1+\frac{2E_2+2E_3}{E{-}E_{\alpha}{+}i\epsilon}+ \frac{4E_2E_3}{\left(E{-}E_{\alpha}{+}i\epsilon\right)^2}\right),
\ee
where notably a factor of 4 instead of 8 appears in the last term, due to the fact that, for \reef{2L2Rforward2}, there is one single time ordering that contributes to the forward limit instead of two, specifically the one in which $t_{1\,{\rm meets}\,3}<t_{2\,{\rm meets}\,3}<t_{2\,{\rm meets}\,4}$.

Due to the different topology, windings now take a slightly different form with respect to \reef{flllr}, and we find
\bea
f_{\reef{2L2Rforward1}}=-\frac{32\ell_s^2}{2!^2}\left(\prod_{i=1}^4\int \frac{\dd E_in_iE_i }{4\pi}\right) \left[ 1-\beta(n_1 e^{\beta E_1}E_1{+}n_3 e^{\beta E_3}E_3)+\frac{4}{3}\beta^2 n_1n_3e^{\beta(E_1+E_3)}E_1E_3\right]
\eea
with a completely analogous expression for $f_{\reef{2L2Rforward2}}$, with the factor of $\frac{4}{3}$ replaced by $\frac{2}{3}$. Notice that the two terms in the numerator of the single $E{-}E_{\alpha}{+}i\epsilon$ denominator of \reef{2L2Rforward1} get a different winding factor, reflecting the different topologies that emerge when either the $P$ or $Q$ propagator collapses into a contact term.

Performing the integral, we find
\begin{align}
f_{\reef{2L2Rforward1}}=-\frac{\ell_s^6\pi^4}{10368\beta^4}\left(\frac{1}{4}-1+\frac{4}{3}\right) ~~~,~~~
f_{\reef{2L2Rforward2}}=-\frac{\ell_s^6\pi^4}{10368\beta^4}\left(\frac{1}{4}-1+\frac{2}{3}\right).
\end{align}
To get to the total yield of the $llrr$ topology, we have to sum the above two contributions, multiply the result by two in order to account for the third and fourth diagrams on the RHS of \reef{2l2rFeyn}, and add the contribution from the covariant contact term of \reef{2l2rFeyn}. This last simply gives
\be\label{contact8}
f_{\rm contact}=-\frac{1}{2!^2}\left(\prod_{i=1}^4\int \frac{\dd E_i n_i}{4\pi E_i}\right)\big(\!-32\ell_s^6 E_1^2E_2^2E_3^2E_4^2\, \big)=+\ \frac{1}{4}\ \frac{\ell_s^6\pi^4}{10368\beta^4}\,.
\ee
Summing everything, we finally find
\be\label{4to4trace}
\begin{tikzpicture}[line width=1.1 pt, scale=.33, baseline=(current bounding box.center)]

		\draw[purple] (-1,-2) -- (2,1) ;
		\draw[purple] (-2,-1) -- (1,2) ;
		\draw[cyan] (1,-2) -- (-2,1) ;
		\draw[white,line width=4] (2,-1) -- (-1,2) ;
		\draw[cyan] (2,-1) -- (-1,2) ;
		\draw[black,fill=gray!15] (0,-1) circle (.2 cm) ;
		\draw[black,fill=gray!15] (1,0) circle (.2 cm) ;
		\draw[black,fill=gray!15] (-1,0) circle (.2 cm) ;
		\node at (-1.3,-2.3) {\scriptsize 3} ;
		\node at (-2.3,-1.3) {\scriptsize 4} ;
		\node at (1.3,-2.3) {\scriptsize 1} ;
		\node at (2.3,-1.3) {\scriptsize 2} ;
		\node at (-1.3,2.3) {\scriptsize 2} ;
		\node at (-2.3,1.3) {\scriptsize 1} ;
		\node at (1.3,2.3) {\scriptsize 4} ;
		\node at (2.3,1.3) {\scriptsize 3} ;
\end{tikzpicture}=\frac{3}{4}\,f_{lllr}
\ee
where the `missing' $\frac{1}{4}$ can be ascribed to the opposite-sign contribution from the covariant contact term.
Given that there are 4 diagrams like the one in \reef{4to4Feyn}, differing only by the location of the interaction vertices, we get that overall
$f_{llrr}=3 f_{lllr}$. Doing the grand total of NNLO effects, we then find
\be
f_{\rm NNLO}=f_{llrr}+f_{lllr}+f_{lrrr}=-\frac{5\ell_s^6\pi^4}{10368\beta^6}
\ee
finding perfect agreement with the square root formula \reef{Esqrt}.

Let us comment on the generalization to arbitrary $D$. The key observation is that, despite the massaging required to bring the amplitude in the OF form, our starting point was always the matrix element $M_{n\to n}$, so we can use at our advantage all the constraints on scattering amplitudes coming from integrability. In particular, at each vertex the only effect of interactions is a flavor-independent phase shift. This means that, when we consider the maximally divergent diagrams to extract the free energy, to account for the different flavors we just need to multiply by $(D-2)^n$, where $n$ is the number of particles partaking the scattering. This is in agreement with the known result.

\subsection{Comparison to TBA and Catalan numbers}

The square root formula \reef{Esqrt} was derived with TBA~\cite{Dubovsky:2012wk}. The basic equations of TBA can be written for a massless theory as (here $\beta=1$)
\bea
f=f_l+f_r&=\sum_{i=l,r} \frac{1}{2\pi}\int_0^\infty \dd E_i \ln \left(1 - e^{-\varepsilon_i(E_i)}\right) \label{freeTBA} \\
\varepsilon_l(E_l)&= E_l+\frac{1}{2\pi}\int_0^\infty \dd E_r \,\partial_{r}\phi(E_l,E_r)\,\ln \left(1 - e^{-\varepsilon_r(E_r)}\right) \label{pseudoTBA}
\eea
with a symmetric expression for $\varepsilon_r(E_r)$. The function $\phi$ that appears in the expression for the `pseudo-energy' $\varepsilon$ is the phase that characterizes the $2\to 2$ scattering in \reef{Sexact}, i.e. $\phi=\ell_s^2 E_lE_r$ for the Flux Tube. While the TBA equations can be solved exactly for this theory, leading to \reef{Esqrt}, we want here to solve them perturbatively by expanding \reef{freeTBA} in powers of $\ell_s^2$, or $\phi$, and interpret them diagramatically.

Given the $l{-}r$ symmetry, we focus on $f_l$ and multiply by 2 the result. At zeroth order we have the usual free-theory free energy
\be
f_l^{(0)}=\frac{1}{2\pi}\int_0^\infty \dd E_l \ln \left(1 - e^{-E_l}\right)=-\frac{\pi}{12}\,.
\ee
When moving to the next orders, we get from Taylor-expanding the logarithm
\be\label{TBA_nlo}
f_l=\frac{1}{2\pi}\int_0^\infty \dd E_l \left( n(E_l)(\varepsilon_l(E_l)-E_l)+\frac{1}{2}n'(E_l)(\varepsilon_l(E_l)-E_l)^2 \right) +\ldots
\ee
where the Bose-Einstein densities appear because $\partial_E \ln(1-e^{-E})=n(E)$, and we kept all terms relevant to go to order $\ell_s^4$. At $O(\ell_s^2)$ we find
\be\label{TBA_lo}
f_l^{(1)}=-\frac{1}{(2\pi)^2}\int_0^\infty \dd E_l \int_0^\infty \dd E_r \ \phi(E_l,E_r) \,n(E_l)n(E_r)=-\frac{\ell_s^2\pi^2}{144}
\ee
where we did a partial integration in $E_r$ to bring $\varepsilon_l-E_l$ in the form of an integral over the Bose-Einstein density $n(E_r)$.

To go to the next order there are now two possibilities: either we take the term proportional to $(\varepsilon_l-E_l)^2$ in \reef{TBA_nlo} and evaluate $\varepsilon_l-E_l$ at leading order, or we take the term proportional to a single power of $\varepsilon_l-E_l$ and evaluate this one at next-to-leading order. We can denote the two terms as respectively $l(r,r)$ and $l(r(l))$. 
Equivalently we can represent diagrammatically the two  possible contributions by
\be
\begin{tikzpicture}[line width=1.1 pt, scale=.7, baseline=(current bounding box.center)]
 		\draw[purple] (.6,-.6) -- (-.6,.6) ;
		\draw[dotted] (-.3,-.7) -- (.2,-.2) ;
		\draw[cyan] (.2,-.2) -- (.7,.3) ;
		\draw[dotted] (-.7,-.3) -- (-.2,.2) ;
		\draw[cyan] (-.2,.2) -- (.3,.7) ;
		\draw[black,fill=gray!15] (-.2,+.2) circle (.1 cm) ;
		\draw[black,fill=gray!15] (.2,-.2) circle (.1 cm) ;
		\node at (.75,-.65) {\scriptsize $l$} ;
		\node at (.9,.4) {\scriptsize $r$} ;
		\node at (.5,.8) {\scriptsize $r$} ;
 \end{tikzpicture} 
 \quad \& \quad 
 \begin{tikzpicture}[line width=1.1 pt, scale=.7, baseline=(current bounding box.center)]
 		\draw[purple] (.6,-.6) -- (-.45,.45) ;
                  \draw[dotted] (1,-.2) -- (.6,.2) ;
                  \draw[purple] (.6,.2) -- (.2,.6) ;
		\draw[dotted] (-.3,-.7) -- (.2,-.2) ;
		\draw[cyan] (.2,-.2) -- (1,.6) ;
		\draw[black,fill=gray!15] (.6,+.2) circle (.1 cm) ;
		\draw[black,fill=gray!15] (.2,-.2) circle (.1 cm) ;
		\node at (.75,-.65) {\scriptsize $l$} ;
		\node at (.05,.7) {\scriptsize $l$} ;
		\node at (1.2,.7) {\scriptsize $r$} ;
		 \end{tikzpicture}  \, , 
\ee
respectively,  which are  full binary trees if the dotted black line is ommited.
Similarly, the LO contribution in \reef{TBA_lo} would read with this notation $l(r)$. 

At NNLO we have five contributions:
$l(r,r,r)$, $l(r(l),r)$, $l(r,r(l))$, $l(r(l,l))$ and $l(r(l(r)))$, and so on. Each term in the expansion comes from either $f_l$ or $f_r$, which determines the `head' of the brackets, and one can always go higher in the Taylor expansion of \reef{TBA_nlo} or deeper in the recursive relation \reef{pseudoTBA}. In the first case we stay at the same level of the brackets and just add more letters, in the second we open new parentheses. Since the $l$ pseudo energy is an integral over the $r$ one and vice versa, every time we go a level deeper we should change the letter.

It can be proven that {\bf (1)} the number of distinct brackets with $n{+}1$ letters is equal to $C_n$, the $n$th Catalan number (or, equivalently, the number of full binary trees with $n$ mothers are counted by $C_n$); {\bf (2)} in the Flux Tube theory, with $\phi=\ell_s^2E_lE_r$, each bracket corresponds to an integral whose value $I_n$ depends just on the order of the expansion, and is given by $I_n=-\ell_s^{2n}({\pi}/{12})^{n+1}$. 
Putting these two pieces of information together, we conclude that
\be
f=-2 \sum_{n=0}^\infty C_n \ell_s^{2n}\left(\frac{\pi}{12}\right)^{n+1}\, ,
\ee
in agreement with the square root formula \reef{Esqrt},
which counts distinct ways of bracketing  a word formed out  of  `$l$' and `$r$' characters or, equivalently, the number of full binary trees. 

It is surely tempting to look for parallels among the TBA and DMB organization of the perturbative expansion of $f$. If in the DMB expansion we focus on the maximally singular diagrams of Fig.~\ref{fig:tubeofflux} --- which are arguably the only ones that contribute to $f$ at any order ---, we find they are organized as phase space integrals weighted by bosonic densities $n(E)$, or their derivatives $n^{(k)}$, times interaction vertices, exactly as for the TBA expansion.

Although this is a remarkable similarity, it is hard to make a full parallel, because already at $O(\ell_s^6)$ we cannot establish a one-to-one correspondence among diagrams. If on one side the DMB expansion is interpretable in terms of tree-level scattering amplitudes with increasing number of legs, the TBA expansion is more suggestive of a decay or branching interpretation, with all possible $1\to n{+}1$ decays at $O(\ell_s^{2n})$. For example, $l(r,r)$ represents a left-moving particle that emits at two consecutive events a righ-mover, while $l(r(l))$ is a left-mover that emits a right-mover which in turn emits a left-mover. The so implied arrow of time is incompatible with an elastic scattering interpretation. At $O(\ell_s^6)$ there are 6 maximally singular forward diagrams instead of 5 (times 2) branching diagrams. Out of the 6 scattering diagrams, 4 are worth $\frac{3}{4}$ of the remaining two, so the 3rd Catalan number $C_3=5$ is reproduced less directly than with the branching interpretation, as $5=2+\frac{3}{4}\times4$. Eq.~\reef{contact8} seems to suggest that it is the $\partial X^8$ vertex which is responsible for ``subtracting the 1 from 6''. At higher orders, higher-order vertices enter. For example, at $O(\ell_s^{10})$ we expect the $\partial X^{12}$ vertex to play a role in the $3l3r\to 3l3r$ amplitude.
It could be interesting    to find a deeper reason for the  appearance of the Catalan numbers.
For example,  why does the integrable amplitude involving a left- and a right-moving intermediate state require a correcting contact term, like in \reef{4to4Feyn}, while the one with two consecutive left-movers does not require it, like in \reef{3l3rperm}? 
These questions are left for future work.

\subsection{Non-universal corrections}
\label{sec:nonuniversal}
In the treatment of the previous sections we made abundant use of the special properties of the integrable realization of the Flux Tube theory, especially to write down amplitudes and streamline computations. However there is no fundamental reason to invoke those properties when using \reef{master}. In the realm of 1+1 dimensional theories, the DMB formula is the perfect tool to study deformations away from integrability. These corrections have to be taken into account in order to match the expected deviation from the square root formula \reef{Esqrt} of  QCD Flux Tubes.

Here we consider the effect of the leading operator deviating from the NG action, that in the static gauge at $D=3$ starts with $L_3=2\gamma_3 (\partial_\mu\partial_\nu X \partial^\mu\partial^\nu X)^2$. We are going to consider effects at $O(\gamma_3)$ and $O(\ell_s^2\gamma_3)$. It turns out that, up to the order we are interested in, the effect of this deformation can be encoded in the phase shift $\phi$ of the $2\to 2$ amplitude, i.e. $\phi(s)=\ell_s^2 s/4+\gamma_3 s^3$.~\footnote{Integrability breaking effects, due to $\gamma_3$, on the vacuum energy appear at higher orders. }

The leading effect is easily computed by mimicking the procedure of Section \ref{sec:LO}, with $M(s)=\ell_s^2 s^2/2+2\gamma_3 s^4+O(s^5)$, and we find $f|_{O(\gamma_3)}=-32\pi^6\gamma_3/225 \beta^8$, in  agreement with \cite{EliasMiro:2019kyf}. 

The next-to-leading effects induced by $\gamma_3$ can be computed using \reef{regNLO}, keeping in mind it requires special treatment for singular contributions to the $3\to 3$ amplitude. The $2\to 2$ amplitude has no effect at $O(\ell_s^2\gamma_3)$ on the partition function  because  it is purely imaginary. The $llr\to llr$ amplitude is given by
\be
M=2048\pi i\,\ell_s^2\gamma_3\left(E_1^2E_2^4E_3^5+E_1^4E_2^2E_3^5\right) [\,\delta(E_1-E_4)+\delta(E_1-E_5)\,]
\ee
and we see that the regular contribution in the forward limit, i.e. the one proportional to $\delta(E_1-E_5)$, gives no yield to \reef{regNLO} because  it is purely imaginary. Let us briefly report the main steps to treat the singular part. Since the $\gamma_3$ operator mediates interactions that are quadratic in each momentum entering the vertex, the numerator of the  covariant amplitude  is proportional to three powers of the propagator $Q$ instead of two (cf. with \reef{covtoOF}). Specifically we have
\be\label{gamma3sing}
M^{(\rm sing)}
=-16\ell_s^2\gamma_3\,\big( \,p_3{\cdot}p_1\ p_4^\mu+p_3{\cdot}p_4\ p_1^\mu\, \big)\,\frac{\,Q_\mu Q_\nu Q_\rho}{Q^2{+}i\epsilon}\,\left(\,(p_6{\cdot}p_2)^2\ p_5^\nu p_5^\rho+(p_6{\cdot}p_5)^2\ p_2^\nu p_2^\rho\,\right)+\ldots.\
\ee
where the ellipses stand for three other Feynman diagrams coming from either changing the location of the $\gamma_3$ vertex or the order of the injection of the pair of legs 1-4 and 2-5. All four terms give the same contribution. 
To translate it to the OF expression, we need the following identity
\be
\frac{\,Q_\mu Q_\nu Q_\rho}{Q^2{+}i\epsilon}=\delta_\mu^0\delta_\nu^0\delta_\rho^0\, Q_0+\left(\delta_\mu^0\delta_\nu^0\delta_\rho^i+\delta_\mu^0\delta_\nu^i\delta_\rho^0+\delta_\mu^i\delta_\nu^0\delta_\rho^0\right)q_i+\frac{1}{2|\vec{Q}|}\left(\,\frac{q_\mu\,q_\nu\,q_\rho}{Q_0{-}|\vec{Q}|{+}i\epsilon}+\frac{\bar{q}_\mu\,\bar{q}_\nu\,\bar{q}_\rho}{{-}Q_0{-}|\vec{Q}|{+}i\epsilon}\,\right)
\ee
with the notation as in \reef{covtoOF}. Like before, the last term corresponds  to an anti-time-ordered OF graph and gives zero in the forward limit. Next  we contract with the external legs momenta, promote to the $E$-dependent amplitude, and take the forward limit, to find
\be\label{non-uni}
\lim_{\rm forward}M^{(\rm sing)}(E)=-4096\,\ell_s^2\gamma_3 \,E_1^2E_2^4E_3^4\left( 1 + \frac{E_3}{E{-}E_\alpha{+}i\epsilon} \right)+(1\leftrightarrow 2)\,.
\ee
Thus, all in all, the contribution to the free energy is given by
\be
f|_{O(\ell_s^2\gamma_3)}=4\times 4096 \, \ell_s^2\gamma_3 \,\frac{1}{2!}\left(\prod_{i=1}^3\frac{\dd E_i n_i}{4\pi E_i}\right) E_1^2 E_2^4E_3^4\left( 1-\frac{\beta}{2}n_3e^{\beta E_3} E_3\right)=-\frac{64\ell_s^2\gamma_3 \pi^7}{675\beta^{10}}
\ee
where the factor 4 in front accounts for both the identical $(1\leftrightarrow 2)$ contribution in \reef{non-uni} and the $lrr\to lrr$ symmetric amplitude. This result agrees with the TBA calculation presented in 
appendix E of \cite{EliasMiro:2019kyf}.

\section{Conclusions}\label{sec:conclusion}
In this work we have developed further  the ideas of Dashen, Ma and Bernstein~\cite{Dashen:1969ep,dashen1970singular} connecting the free energy of a thermodynamic system to its $S$-matrix elements. From a technical perspective, 
 the most challenging task we have addressed is resolving   the apparent IR divergences that affect this formalism, which stem from the evaluation of amplitudes in a forward configuration. As a concrete benchmark we picked the theory of a Flux Tube embedded in $D$ dimensions and described, at the lowest order, by the Nambu Goto action. This choice has two main justifications. Firstly, this system allows for some striking simplifications in specific settings: notably the assumption of integrability greatly simplifies the structure of the $S$-matrix, and allows the   calculation of   the free energy at all-orders in the perturbative expansion. This provided an excellent testing ground for the formalism.  In this work, we successfully reproduced  Eq.~\eqref{Esqrt} up to and including the NNLO perturbative corrections, 
 demonstrating a general method for handling the aforementioned IR divergences, which can be applied to other theories in any dimension. 
 
The second justification for studying the Flux Tube EFT lies in its connection to Yang-Mills theory, as it can be used to describe the chromoelectric flux in the confining phase. For a realistic description of Yang-Mills confining flux tubes, the Nambu-Goto action must be refined with  higher-dimensional operators that break integrability. 
 Despite this added complexity, it does not introduce technical difficulties or new theoretical challenges to the formalism presented here. For instance, in Sec.~\ref{sec:nonuniversal}, we demonstrated the impact of the first non-universal contribution to the flux tube ground state energy.

This work leaves several possible avenues to be explored. In Section~\ref{sec:QCD} we showed how the present method can be used to compute the partition function of QCD up to $O(\alpha_s)$ in perturbation theory. Famously, perturbative computations 
of the QCD equation of state are   plagued with IR divergences, associated to either soft collective excitations or with collinear scatterings (see e.g. \cite{Andersen:2004fp,Ghiglieri:2020dpq} for a review). It would be very  interesting to investigate whether the DMB formula can provide a new perspective on how to treat these IR divergences when studying the QCD equation of state at higher-orders. 
In particular,  at $O(\alpha_s^2)$, the canonical treatment is to introduce a Debye mass. In the DMB formalism it is more natural to proceed with other regulators of IR safe collider observables.  
 At $O(\alpha_s^3)$ the IR divergences are of  non-perturbative nature, the dimensional reduction of     QCD 
 to magnetostatic   QCD is IR finite due to the non-perturbative mass gap. In perturbation theory these   effects first appear in IR divergent  four-loop thermal vacuum diagrams~\cite{Linde:1980ts}.~\footnote{In this work, we have meticulously carried out the Flux Tube calculations up to the same four-loop order.}
 
The on-shell formalism employed in the present work for studying the system in a thermal bath admits   generalizations that we did not explore here, like the coupling to  external fields. Consider adding for example a magnetic field $h$ via a coupling $\int h(x)O(x)$. Under appropriate assumptions we can still define a ($E$-dependent) scattering matrix $S_{[h]}(E)$, for which the DMB formula \reef{master} is expected to hold. By taking a derivative with respect to the external field of \reef{master}, we get
\be\label{response}
\frac{\delta F[h]}{\delta h (x)}=-\frac{1}{2\pi i}\int \dd E e^{-\beta E} \Tr_c \left[S^{-1}_{[h]}(E)\frac{\delta S_{[h]}(E)}{\delta h (x)}\right]
\ee
where the LHS is equal to the local magnetization $m(x)$ when $h\to 0$, whereas in the same limit we have on the RHS that $\langle \beta | \delta S / \delta h (x) |\alpha \rangle$ can be identified with the form factor $i O_{\beta\alpha}(x)$ \cite{Weinberg:1995mt} of the operator coupled to the magnetic field. Formula \reef{response} for the local magnetization, which to our knowledge is new, can be obviously generalized to an arbitrary number of external fields and $\delta/\delta h(x)$ variations. Moreover, it is structurally completely analogous to the master formula, so it admits a similar diagrammatic interpretation, all the comments on the spurious forward divergences are expected to carry over, etc. When we consider the (Fourier transform of the) second variational derivative of $F$ with respect to some external field, we can extract information of the thermal dispersion relation of particles excited by $O(x)$, the operator coupled to the external field. This approach could also unify in a single formalism 
previous results 
 connecting thermal contribution to the mass with $S$-matrix elements \cite{Jeon:1998zj}.
Further study of \reef{response} and of excited states following the above discussion is left for future work.

Recently there has been a flourishing of methods to constrain  $S$-matrix elements without referring to a Lagrangian, for example via $S$-matrix bootstrap, implementing causality constraints or with  geometric construction, see e.g. \cite{Arkani-Hamed:2020blm, Caron-Huot:2020cmc, Tolley:2020gtv, Bellazzini:2020cot, Paulos:2017fhb} and references therein. The results here presented   could be useful to compute the partition function of theories characterised  by $S$-matrices leaving at the boundaries of the allowed spaces found in these works. This could provide a renewed  interesting interplay between constraints from thermal physics considerations (e.g. speed of sound boundedness) 
and constraints from causality and unitarity of  physics in the vacuum. 

\paragraph{Note added:}
As this work was nearing completion (see \cite{miro:IFAE} for a seminar on the topic), ref.~\cite{Schubring:2024yfi} appeared on arXiv, presenting some overlapping considerations and conclusions. This reference demonstrates  the application  of the  DMB formula in  $d=1+1$ massive  integrable theories.

\section*{Acknowledgements}
We are grateful to J.R.~Espinosa, A.~Guerrieri, A.~Pomarol and G.~Villadoro for useful discussions. 
PB is supported by the Slovenian Research Agency under the research core funding No. P1-0035 and in part by the research grants N1-0253 and J1-4389. JEM is supported by the European Research Council, grant agreement n. 101039756.
EG is supported by the Collaborative Research Center SFB1258 and the Excellence Cluster ORIGINS, which is funded by the Deutsche Forschungsgemeinschaft (DFG, German Research Foundation) under Germany’s Excellence Strategy – EXC-2094-390783311.



\appendix

\section{Two-level system}\label{app_2level}

The derivation of the master formula \reef{master} is completely general, therefore it can be checked against one of the simplest  quantum-mechanical model, a two-level system with Hamiltonian
\begin{align}\label{2levels}
	H &= H_0 + V & H_0 &=\begin{pmatrix}
		E_1&0\\
		0 &E_2
	\end{pmatrix} & V&=\begin{pmatrix}
		0& v\\
		v &0
	\end{pmatrix}
\end{align}
where we assume $E_1-E_2\gg v>0$. 
The partition function can be computed exactly:
\begin{align}
	{Z}=\Tr\, e^{-\beta H}&=2\,e^{-\frac{\beta(E_1+E_2)}{2}}\cosh \frac{\beta {\Delta}}{2}
	&\text{with~~} \Delta=\sqrt{(E_1-E_2)^2+4 v^2}\ .
	\label{eq:exactexpression}
\end{align}
Expanded to $\order{v^4}$ the previous expression becomes
\begin{align}
	{Z}=e^{-\beta E_1}+e^{-\beta E_2}-\beta \,v^2\,\frac{e^{-\beta E_1}-e^{-\beta E_2}}{E_1-E_2} +\beta^2 v^4 \left(\frac{ e^{-\beta E_1}+e^{-\beta E_2}}{2 (E_1-E_2)^2}+\frac{e^{-\beta E_1}-e^{-\beta E_2}}{\beta (E_1-E_2)^3}\right),
	\label{eq:order4}
\end{align}
where we can already recognize the zeroth order contribution to be ${Z}_0$.

We want to reproduce the interacting part if \eq{eq:order4} using the master formula \eq{masterformula}. 
In this case we do not have a notion of asymptotic states and a  scattering process among them. 
However  the operator $S(E)$ is still defined in terms of resolvents, which are now 2 by 2 matrices. The free-theory one reads
\begin{align}
	G_0(z)&=(z-H_0)^{-1}=\begin{pmatrix}
		\frac{1}{z-E_1} & 0\\
		0 &\frac{1}{z-E_2} 
	\end{pmatrix} .
	\label{eq:order0resolvent}
\end{align}
Eq.~\eqref{eq:order0resolvent} allows us to define 
$\delta(E-H_0)\equiv \left(G_0(E-i\epsilon)-G_0(E+i\epsilon)\right)/(2\pi i)$.
The next step is to identify all relevant contributions to $\Delta {Z}={Z}-{Z}_0$ at the desired order. First of all we expand the transfer matrix $T(z)$ in $v$ up to $\order{v^4}$ using the Lippmann-Schwinger formula \eq{T_OFPT} --- which says that $T^{(n)}=V(G_0V)^{n-1}$ ---, and find $T^{(1)}(z)=V$, while
\begin{align}
	T^{(2)}(z)=v^2 \begin{pmatrix}
		\frac{1}{z-E_2} & 0\\
		0 &\frac{1}{z-E_1} 
	\end{pmatrix} 
	& ~~~,~~~~~~~T^{(3)}(z)=v^3\left(
	\begin{array}{cc}
		0 & \frac{1}{(z-E_1) (z-E_2 )} \\
		\frac{1}{(z-E_1) (z-E_2 )} & 0 \\
	\end{array}
	\right)~~~,\nonumber \\
	T^{(4)}(z)&=v^4 \begin{pmatrix}
		\frac{1}{(z-E_1) (z-E_2)^2} & 0\\
		0 &\frac{1}{(z-E_1)^2 (z-E_2)}
	\end{pmatrix}\,.
\end{align}
At $\order{v}$ we can easily see that there is no contribution, as expected from \eq{eq:order4}. Indeed at this order \eq{master} reads
$
\Delta {Z}^{(1)}=-\beta\int\dd E e^{-\beta E}\Tr(\delta(E-H_0)V)
$,
which is zero because $\delta(E-H_0)V$ has only zeros on the diagonal so its trace vanishes. With the same reasoning one can show that all the odd orders of \eq{masterformula} are zero for the Hamiltonian in \eq{2levels}.

Moving to the next orders, we find it useful to introduce the shorthand notation $G_0\equiv G_0(E+i\epsilon)$ and $\bar{G}_0\equiv G_0(E-i\epsilon)$. 
As it is more instructive, we move on to directly computing the $\order{v^4}$ contribution.
Calling now $\Delta G_0 \equiv \bar{G}_0-G_0$, we find from expanding $\ln S(E)$ to this order several terms
\begin{align}
		\label{eq:freeenergy2statesorder4}
	\Delta {Z}^{(4)}=-\frac{\beta}{2\pi i} \int \dd E\,e^{-\beta E}  \Tr\, &[\, \Delta G_0 \,T^{(4)}+\Delta G_0 \, T^{(1)} \Delta G_0 \,T^{(3)} \\
	&+\frac{1}{2}\Delta G_0 \,T^{(2)}\Delta G_0\,T^{(2)}+\left(\Delta G_0 \,T^{(1)}\right)^2 \Delta G_0 \,T^{(2)}+\frac{1}{4}
	\left(\Delta G_0  \,T^{(1)}\right)^4\, ]\,. \nonumber
\end{align}
To simplify the sum of all five contributions we make use of the following identities
\begin{align}
	(G_0)^{n+1}&=\frac{(-1)^{n}}{n!}\,\partial_E^n G_0\\
	G_{0}\,\bar{G}_0&=\frac{1}{2}\, G_0|_{\epsilon=0}\left(G_0+\bar{G}_0\right)\ ,
\end{align}
which are true in general, as well as $
G_0V+VG_0= V \Tr G_0$, valid in our simple problem. 
Then, as can be verified explicitly, the sum can be rewritten as a linear combination of $\delta(E-H_0)$ and its first derivative, as follows
\begin{align}
	\Delta {Z}^{(4)}=-\frac{\beta v^4}{2\pi i}\int\dd E \,e^{-\beta E} \Tr[\,\frac{1}{2}M^2\,\partial_E (\bar{G}_0-G_0)+M^3 (\bar{G}_0-G_0 )]\ ,
	\label{eq:deltaFsecondorder}
\end{align}
where we also defined the $E$-independent matrix $M\equiv\text{diag}[(E_1-E_2)^{-1},(E_2-E_1)^{-1} ]$.

In Eq.~\eqref{eq:deltaFsecondorder} we can integrate by parts the derivative acting on the delta and obtain
\begin{align}
	\Delta {Z}^{(4)}=\beta^2 v^4 \left(\frac{ e^{-\beta E_1}+e^{-\beta E_2}}{2 (E_1-E_2)^2}+\frac{e^{-\beta E_1}-e^{-\beta E_2}}{\beta(E_1-E_2)^3}\right)
\end{align}
matching with the $\order{v^4}$ term of Eq.~\eqref{eq:order4}.
Here we see in a much simpler setting an interesting phenomenon that we also encounter in the main text: all terms in Eq.~\eqref{eq:freeenergy2statesorder4} but the first one, taken individually, have a divergence caused by factors of $\Delta G_0\propto \delta(E-H_0)$. However this is cured in the final expression, where all contributions sum to give a derivative of the delta distribution. The same happens in Section~\ref{sec:fluxtuberecap}, where many contributions, taken on their own, present singularities that are canceled when all pieces are carefully taken into account.

\section{Old Fashioned and covariant perturbation theory}

We presented \reef{covtoOF} as an identity, and showed that it had the property of providing an OFPT decomposition of \reef{3to3sing}. Here we would like to proceed in the opposite direction. We start from a toy derivatively coupled Lagrangian
\be\label{toyL}
L=L_0[\phi,\phi',\varphi]+c\, \partial_\mu \varphi\,\partial^\mu \phi \,(\partial_\nu\phi)^2+c' \,\partial_\mu \varphi \,\partial^\mu \phi'\, (\partial_\nu\phi')^2\,.
\ee
There is only one tree-level diagram contributing to the process $3\phi\to3\phi'$, mediated by a $\varphi$ exchange. It is proportional to $cc'$ and has amplitude
\be\label{m33cov}
M(\phi\,\phi\,\phi\to\phi'\phi'\phi')=4cc'\,A_\mu \,\frac{Q^\mu Q^\nu}{Q^2+i\epsilon}\,B_\nu\,,
\ee
where $Q=p_1{+}p_2{+}p_3$, while $A=s_{12}p_3{+}s_{13}p_2{+}s_{23}p_1$, $B=s_{45}p_6{+}s_{46}p_5{+}s_{56}p_4$.

We want now to obtain the same tree-level amplitude using OFPT. First of all we need to calculate the interaction Hamiltonian $V=\int \dd^d x \,v(x)$, where $v(x)$ is not just $-L(x)$ due to the derivative interactions. To obtain the Hamiltonian one needs to perform a Legendre transform of the Lagrangian with respect to $\dot{\phi},\dot{\phi}'$ and $\dot{\varphi}$. Truncating the transform at $O(c^2,c'^2)$, we find after some straightforward but boring algebra
\be
v(x)=-c(\pi^2{-}\nabla\phi^2)(\pi \varpi{-}\nabla \phi\nabla\varphi)-c'(\pi'^2{-}\nabla\phi'^2)(\pi' \varpi{-}\nabla \phi'\nabla\varphi)+cc'\pi\pi'(\pi^2{-}\nabla\phi^2)(\pi'^2{-}\nabla\phi'^2)
\ee
where $\pi,\pi'$ and $\varpi$ are the conjugate momenta to respectively $\phi,\phi'$ and $\varphi$. While the terms proportional to $c$ and $c'$ are just the negative of the covariant Lagrangian interactions of \reef{toyL}, the non-covariant term proportional to $cc'$ is new.

In OFPT there are now three contributions to consider at order $cc'$, given by
\be
M(\phi\,\phi\,\phi\to\phi'\phi'\phi')~~=~~~~
\begin{tikzpicture}[line width=.9 pt, scale=.33, baseline=(current bounding box.center)]

	\draw (0,-2) -- (0,2) ;
	\draw (-1,-2) -- (0,-1) -- (1,-2) ;
	\draw (-1,2) -- (0,1) -- (1,2) ;
	\node at (.6,-.9) {\scriptsize $c$} ;
	\node at (.7,1) {\scriptsize $c'$} ;
	
	\begin{scope}[shift={(6,0)}]
	\draw (0,-2) -- (0,1) -- (2,-1) -- (2,2) ;
	\draw (-1,-2) -- (0,1) -- (1,-2) ;
	\draw (1,2) -- (2,-1) -- (3,2) ;
	\draw[white,line width=1.6] (.3,1.15) -- (-.3,1.0);
	\draw[white,line width=1.6] (2.3,-1.) -- (1.7,-1.15);
	\node at (0,1.4) {\scriptsize $c$} ;
	\node at (2.2,-1.3) {\scriptsize $c'$} ;
	\end{scope}
	
	\begin{scope}[shift={(14,0)}]
	\draw (0,-2) -- (0,2) ;
	\draw (-1,-2) -- (1,2) ;
	\draw (-1,2) -- (1,-2) ;
	\node at (1.1,0.2) {\scriptsize $cc'$} ;
	\end{scope}
	
	\draw[white] (0,-2) -- (0,-2.4) ;
	
	\node at (3,0) {+};
	\node at (11,0) {+};
	
\end{tikzpicture}
\ee
where the third diagram comes from the non-covariant contact term, while the first two correspond to the two time orderings of the events ``$c$-interaction'' and ``$c'$-interaction''. Altogether the three contributions give
\be
M(\phi\,\phi\,\phi\to\phi'\phi'\phi')=4cc'\,A_\mu \,B_\nu \left( \delta_\mu^0\delta_\nu^0+\frac{q_\mu\,q_\nu}{2|\vec{Q}|(Q_0{-}|\vec{Q}|{+}i\epsilon)}+\frac{\bar{q}_\mu\,\bar{q}_\nu}{2|\vec{Q}|({-}Q_0{-}|\vec{Q}|{+}i\epsilon)}\right)
\ee
where $q_\mu$ and $\bar{q}_\mu$ (see below \reef{covtoOF} for their definition) are on-shell due to the OFPT rules, as opposed to $Q_\mu$ which is off-shell according to the usual four-momentum conservation rule at the vertex. Using \reef{covtoOF} we see that the OFPT amplitude is the same as the covariant one, \reef{m33cov}.

\section{Feynman diagrams and factorization}\label{ls4_appendix}

To resolve the $\delta(0)$ singularities that the integrable amplitudes feature in the forward limit, we decomposed the amplitudes as a sum of propagators with appropriate numerators, like in \reef{3to3Feynman}, \reef{3l3rperm} and \reef{4to4Feyn}.
These can be taken just as mathematical identities. However it is kind of reassuring to check that integrable amplitudes emerge from the usual Feynman rules starting from the Nambu-Goto action. In this appendix we sketch the derivation of $llr\to llr$ and $llrr\to llrr$ amplitudes using Feynman diagrams.

	\subsection*{Factorization of $M(1_l,2_l,3_r \to 4_l,5_l,6_r)$}
	
	We consider the Nambu-Goto theory with one flavor, whose Lagrangian is
	\be
	L=-\frac{1}{\ell^2}\sqrt{1-\ell^2(\partial X)^2}=-\frac{1}{\ell^2}+\frac{1}{2}\partial X^2+\frac{\ell^2}{8}\partial X^4+\frac{\ell^4}{16}\partial X^6+\frac{5\ell^6}{128}\partial X^8+O(\ell^8).
	\ee
	The tree-level contribution to the process $llr\to llr$ is at $O(\ell^4)$ and it comes entirely from two insertions of $\partial X^4$, because the contact topology with $\partial X^6$ vanishes on-shell. Moreover, the L and R particles must be distributed according to the following pattern not to get a trivial zero at the vertex
	\begin{center}
	\begin{tikzpicture}[line width=1.1 pt, scale=.7, baseline=(current bounding box.center)]
	
		\draw (0,1) to  (3,1) ;
		\draw (1,0) to  (1,2) ;
		\draw (2,0) to  (2,2) ;

		\node at (-.5,1) {$R$} ;
		\node at (3.5,1) {$R$} ;
		
		\node at (1,-0.5) {$L$} ;
		\node at (2,-0.5) {$L$} ;;
		
		\node at (1,2.5) {$L$} ;
		\node at (2,2.5) {$L$} ;

	 \end{tikzpicture}
	 \end{center}
There are {\bf\textcolor{blue}{6}} distinct Feynman diagrams. Each of them can be uniquely defined by a (redundant) expression like $M(12|45)$, specifying that particles $1_l,2_l$ join the $3_r$ vertex (whereas $4_l,5_l$ join $6_r$).
To evaluate the diagram, it is useful to keep in mind that the propagator $q$ can be decomposed as $q_l+q_r$. When evaluating the vertices, that are of the form $(k{\cdot}k)(k{\cdot}k)$, only $q_r$ enters, because there are already two external left-movers entering each vertex. For all diagrams $q_r=p_3$. We find
\be
M(12|45)=-\ell^4\frac{(2\,p_1{\cdot}p_3 \,p_2{\cdot}p_3)(2\,p_4{\cdot}p_3 \,p_5{\cdot}p_3)}{2\, p_3{\cdot}(p_1{+}p_2)+i\epsilon}=-16\ell^4\frac{E_1E_2E_4E_5E_3^3}{E_1{+}E_2{+}i\epsilon}
\ee
When we compute $M(45|12)$, we find it is just \emph{minus} the complex conjugate to $M(12|45)$, that is $M(45|12)=-16\ell^4E_1E_2E_4E_5E_3^3/({-}E_1{-}E_2{+}i\epsilon)$. When we sum up the two, only the imaginary part remains
\be
M(12|45)+M(45|12)=32i\pi\ell^4 \, E_1E_2E_4E_5E_3^3 \,\delta(E_1{+}E_2)
\ee
Given that $E_1$ and $E_2$ are both positive, the Dirac-delta imposes $E_1{=}E{=}0$, so the above sum vanishes. The only way to get a non-zero yield is to have an incoming and an outgoing left-mover joining the same vertex. As one can verify, $M(14|25)+M(25|14)+M(15|24)+M(24|15)$ exactly reproduces \reef{123to456} on the support of the deltas. The key to get the integrable structure is the cancellation of the principal value, or real part of the individual Feynman diagrams.


\subsection*{Factorization of $M(1_l,2_l,3_r,4_r \to 5_l,6_l,7_r,8_r)$}
	
Amplitude $M(1_l,2_l,3_r,4_r \to 5_l,6_l,7_r,8_r)$ has a total of {\bf\textcolor{blue}{280}} 3-vertex Feynman diagrams, {\bf\textcolor{blue}{64}} trivially zero and {\bf\textcolor{blue}{216}} with one of the following patterns
	\begin{center}
	\begin{tikzpicture}[line width=1.1 pt, scale=.7, baseline=(current bounding box.center)]
	
		\draw (0,1) to  (4,1) ;
		\draw (1,0) to  (1,2) ;
		\draw (2,0) to  (2,2) ;
		\draw (3,0) to  (3,2) ;

		\node at (-.5,1) {$L$} ;
		\node at (4.5,1) {$R$} ;
		
		\node at (1,-0.5) {$R$} ;
		\node at (2,-0.5) {$L$} ;
		\node at (3,-0.5) {$L$} ;
		
		\node at (1,2.5) {$R$} ;
		\node at (2,2.5) {$R$} ;
		\node at (3,2.5) {$L$} ;
		
		\node at (2,-1.5) {$({\bf \textcolor{blue}{144}})$} ;
		
		\begin{scope}[shift={(7,0)}]
		
		\draw (0,1) to  (4,1) ;
		\draw (1,0) to  (1,2) ;
		\draw (2,0) to  (2,2) ;
		\draw (3,0) to  (3,2) ;

		\node at (-.5,1) {$L$} ;
		\node at (4.5,1) {$L$} ;
		
		\node at (1,-0.5) {$R$} ;
		\node at (2,-0.5) {$L$} ;
		\node at (3,-0.5) {$R$} ;
		
		\node at (1,2.5) {$R$} ;
		\node at (2,2.5) {$L$} ;
		\node at (3,2.5) {$R$} ;
		
		\node at (2,-1.5) {$({\bf \textcolor{blue}{36}{+}\textcolor{blue}{36}})$} ;
		
		\end{scope}

	 \end{tikzpicture}
	 \end{center}
\textcolor{blue}{\bf 36+36:} Let us start by computing the second class, which is simpler. It is zero when any of the propagators goes on-shell, so we can directly remove the $i\epsilon$. Using the all-incoming convention and that $Q_1{+}Q_2{+}Q_3{+}Q_4=0$, we find
\begin{equation}
	\begin{tikzpicture}[line width=1.1 pt, scale=.5, baseline=(current bounding box.center)]
	
		\draw (0,1) to  (4,1) ;
		\draw (1,0) to  (1,2) ;
		\draw (2,0) to  (2,2) ;
		\draw (3,0) to  (3,2) ;

		\node at (-.5,1) {\footnotesize $P_1$} ;
		\node at (4.5,1) {\footnotesize $P_3$} ;
		
		\node at (1,-0.5) {\footnotesize $Q_1$} ;
		\node at (2,-0.5) {\footnotesize $P_2$} ;
		\node at (3,-0.5) {\footnotesize $Q_2$} ;
		
		\node at (1,2.5) {\footnotesize $Q_4$} ;
		\node at (2,2.5) {\footnotesize $P_4$} ;
		\node at (3,2.5) {\footnotesize $Q_3$} ;
		
	 \end{tikzpicture}=2 \,\frac{P_1{\cdot}Q_1 P_1{\cdot}Q_4{\times}P_2{\cdot}(Q_1{+}Q_4)P_4{\cdot}(Q_2{+}Q_3){\times} P_3{\cdot}Q_2 P_3{\cdot}Q_3}{P_1{\cdot}(Q_1{+}Q_4)P_3{\cdot}(Q_2{+}Q_3)}=32\prod_i E_i\,,
\end{equation}
where the result is valid for any incoming-outgoing configuration of $P$s and $Q$s.
Given that there is a total of 72 diagrams of this kind, each giving the same yield, their grand total is $2304\prod_i E_i$.

\vspace{.3cm}

\noindent \textcolor{blue}{\bf 144:} This structure is the richest. We find
\begin{equation}
	\begin{tikzpicture}[line width=1.1 pt, scale=.7, baseline=(current bounding box.center)]
	
		\draw (0,1) to  (4,1) ;
		\draw (1,0) to  (1,2) ;
		\draw (2,0) to  (2,2) ;
		\draw (3,0) to  (3,2) ;

		\node at (-.5,1) {\footnotesize $P_1$} ;
		\node at (4.5,1) {\footnotesize $Q_1$} ;
		
		\node at (1,-0.5) {\footnotesize $Q_3$} ;
		\node at (2,-0.5) {\footnotesize $Q_2$} ;
		\node at (3,-0.5) {\footnotesize $P_3$} ;
		
		\node at (1,2.5) {\footnotesize $Q_4$} ;
		\node at (2,2.5) {\footnotesize $P_2$} ;
		\node at (3,2.5) {\footnotesize $P_4$} ;
		
	 \end{tikzpicture}=\frac{N_1+N_2}{(Q_1{\cdot}(P_3{+}P_4){+}i\epsilon)(P_1{\cdot}(Q_3{+}Q_4){+}i\epsilon)}
\end{equation}
where
\begin{align}
N_1&=P_1{\cdot}Q_3P_1{\cdot}Q_4{\times}Q_1{\cdot}P_3Q_1{\cdot}P_4{\times}(P_1{\cdot}Q_1P_2{\cdot}Q_2+P_1{\cdot}Q_2P_2{\cdot}Q_1) \\
N_2&=P_1{\cdot}Q_3P_1{\cdot}Q_4{\times}Q_1{\cdot}P_3Q_1{\cdot}P_4{\times}\big(P_2{\cdot}Q_2(P_3{+}P_4){\cdot}(Q_3{+}Q_4){+}P_2{\cdot}(Q_3{+}Q_4)Q_2{\cdot}(P_3{+}P_4)\big)
\end{align}
The denominator produces 4 terms, having one, two or no Dirac $\delta$s. The numerator $N_2$ vanishes on the support of each $\delta$. The numerator $N_1$ vanishes too on the support of each $\delta$, unless the following condition is satisfied: \emph{$P_3$ and $P_4$ must be one in the initial and one in the final state; the same for $Q_3$ and $Q_4$.} This condition singles out {\bf\textcolor{blue}{64}} \# (hash) diagrams.

Given that $N_2$ is linear in the momenta appearing in the denominator, it is the easiest to compute
\be
\frac{N_2}{Q_1{\cdot}(P_3{+}P_4)P_1{\cdot}(Q_3{+}Q_4)}=32 \prod_i E_i\,,
\ee
a result that is independent on the choice of $P$s and $Q$s.

In order to simplify the analogous expression with $N_1$, where the propagators do not cancel with the numerators, it is convenient to sum over the 4 diagrams with the same propagator structure, that is
\be
\begin{tikzpicture}[line width=1.1 pt, scale=.6, baseline=(current bounding box.center)]
	
		\draw (0,1) to  (4,1) ;
		\draw (1,0) to  (1,2) ;
		\draw (2,0) to  (2,2) ;
		\draw (3,0) to  (3,2) ;

		\node at (-.5,1) {\footnotesize $P_1$} ;
		\node at (4.5,1) {\footnotesize $Q_1$} ;
		
		\node at (1,-0.5) {\footnotesize $Q_3$} ;
		\node at (2,-0.5) {\footnotesize $Q_2$} ;
		\node at (3,-0.5) {\footnotesize $P_3$} ;
		
		\node at (1,2.5) {\footnotesize $Q_4$} ;
		\node at (2,2.5) {\footnotesize $P_2$} ;
		\node at (3,2.5) {\footnotesize $P_4$} ;

		\begin{scope}[shift={(6.5,0)}]
		\draw (0,1) to  (4,1) ;
		\draw (1,0) to  (1,2) ;
		\draw (2,0) to  (2,2) ;
		\draw (3,0) to  (3,2) ;

		\node at (-.5,1) {\footnotesize $P_1$} ;
		\node at (4.5,1) {\footnotesize $Q_2$} ;
		
		\node at (1,-0.5) {\footnotesize $Q_3$} ;
		\node at (2,-0.5) {\footnotesize $Q_1$} ;
		\node at (3,-0.5) {\footnotesize $P_3$} ;
		
		\node at (1,2.5) {\footnotesize $Q_4$} ;
		\node at (2,2.5) {\footnotesize $P_2$} ;
		\node at (3,2.5) {\footnotesize $P_4$} ;
		\end{scope}

		\begin{scope}[shift={(13,0)}]
		\draw (0,1) to  (4,1) ;
		\draw (1,0) to  (1,2) ;
		\draw (2,0) to  (2,2) ;
		\draw (3,0) to  (3,2) ;

		\node at (-.5,1) {\footnotesize $P_2$} ;
		\node at (4.5,1) {\footnotesize $Q_1$} ;
		
		\node at (1,-0.5) {\footnotesize $Q_3$} ;
		\node at (2,-0.5) {\footnotesize $Q_2$} ;
		\node at (3,-0.5) {\footnotesize $P_3$} ;
		
		\node at (1,2.5) {\footnotesize $Q_4$} ;
		\node at (2,2.5) {\footnotesize $P_1$} ;
		\node at (3,2.5) {\footnotesize $P_4$} ;
		\end{scope}

		\begin{scope}[shift={(19.5,0)}]
		\draw (0,1) to  (4,1) ;
		\draw (1,0) to  (1,2) ;
		\draw (2,0) to  (2,2) ;
		\draw (3,0) to  (3,2) ;

		\node at (-.5,1) {\footnotesize $P_2$} ;
		\node at (4.5,1) {\footnotesize $Q_2$} ;
		
		\node at (1,-0.5) {\footnotesize $Q_3$} ;
		\node at (2,-0.5) {\footnotesize $Q_1$} ;
		\node at (3,-0.5) {\footnotesize $P_3$} ;
		
		\node at (1,2.5) {\footnotesize $Q_4$} ;
		\node at (2,2.5) {\footnotesize $P_1$} ;
		\node at (3,2.5) {\footnotesize $P_4$} ;
		\end{scope}
		
	 \end{tikzpicture}
\ee
which remarkably gives the universal contact term $32\prod_i E_i$ upon using $P_1{+}P_2{+}P_3{+}P_4=0$ and $Q_1{+}Q_2{+}Q_3{+}Q_4=0$. We conclude that each diagram gives on average a total of $(32+8) \prod_i E_i$, for a grand total of $5760 \prod_i E_i$.

Let us finally consider the $\delta$ structure generated by the \# diagrams. We find
\begin{equation}
	\begin{tikzpicture}[line width=1.1 pt, scale=.7, baseline=(current bounding box.center)]
	
		\draw (0,1) to  (4,1) ;
		\draw (1,0) to  (1,2) ;
		\draw (2,0) to  (2,2) ;
		\draw (3,0) to  (3,2) ;
		
		\draw[red] (1.25,.75) to (1.75,1.25) ;
		\draw[red] (2.25,.75) to (2.75,1.25) ;
		
		\node at (-.5,1) {\footnotesize $P_1$} ;
		\node at (4.5,1) {\footnotesize $Q_1$} ;
		
		\node at (1,-0.5) {\footnotesize $Q_3$} ;
		\node at (2,-0.5) {\footnotesize $Q_2$} ;
		\node at (3,-0.5) {\footnotesize $P_3$} ;
		
		\node at (1,2.5) {\footnotesize $Q_4$} ;
		\node at (2,2.5) {\footnotesize $P_2$} ;
		\node at (3,2.5) {\footnotesize $P_4$} ;
		
	 \end{tikzpicture}=-32\pi^2 E_{P_1}E_{Q_1}\delta(E_{P_3}{-}E_{P_4})\delta(E_{Q_3}{-}E_{Q_4})\prod_i E_i
	 \end{equation}

\vspace{.3cm}

\noindent \textcolor{blue}{\bf 24+24:} In order to compute the full amplitude, we should add diagrams	
	\begin{center}
	\begin{tikzpicture}[line width=1.1 pt, scale=.7, baseline=(current bounding box.center)]
	
		\draw (0,1) to  (4,1) ;
		\draw (1,0) to  (1,2) ;
		\draw (2,0) to  (3,2) ;
		\draw (3,0) to  (2,2) ;

		\node at (-.5,1) {$R$} ;
		\node at (4.5,1) {$L$} ;
		
		\node at (1,-0.5) {$L$} ;
		\node at (2,-0.5) {$L$} ;
		\node at (3,-0.5) {$L$} ;
		
		\node at (1,2.5) {$R$} ;
		\node at (2,2.5) {$R$} ;
		\node at (3,2.5) {$R$} ;

	 \end{tikzpicture}
	 \end{center}
Denoting by $V_{4,6}$ the vertices with respectively 6 and 4 legs, we see that only $q_r$ matters for $V_6$, and only $q_l$ for $V_4$. For example, the following diagram equals
\begin{equation}
	\begin{tikzpicture}[line width=1.1 pt, scale=.7, baseline=(current bounding box.center)]
	
		\draw (0,1) to  (4,1) ;
		\draw (1,0) to  (1,2) ;
		\draw (2,0) to  (3,2) ;
		\draw (3,0) to  (2,2) ;

		\node at (-.5,1) {$3$} ;
		\node at (4.5,1) {$6$} ;
		
		\node at (1,-0.5) {$1$} ;
		\node at (2,-0.5) {$2$} ;
		\node at (3,-0.5) {$5$} ;
		
		\node at (1,2.5) {$4$} ;
		\node at (2,2.5) {$7$} ;
		\node at (3,2.5) {$8$} ;

	 \end{tikzpicture}=-3\,\frac{V_4(p_1,p_1,p_3,p_4)V_6(p_2,p_5,p_6,p_3{+}p_4,p_7,p_8)}{2p_1{\cdot}(p_3{+}p_4){+}i\epsilon} =-{288} \prod_i E_i
\end{equation}
Given that there is a total of 48 diagrams of this kind, each giving the same yield, their grand total amounts to $-13824 \prod_i E_i$.

\vspace{.3cm}

\noindent \textcolor{blue}{\bf 1:}	 We are left with the computation of the simplest Feynman diagram, coming from a single insertion of $\partial X^8$. The grand total is given by the product of (value of single contraction) $\times$ (number of inequivalent contractions) $\times$ (number of equivalent contractions)
\be
\left(\frac{5}{128}\, 2^4 \prod_i E_i \right) \times 4! \times 8!!=5760\prod_i E_i
\ee

\subsubsection*{Vanishing of contact terms}
When we sum up all the contact contributions coming from the various topologies, we find that, quite remarkably, they cancel with each other
\be
5760+5760+2304-13824=2^73^2(5+5+2-12)=0.
\ee
Terms with only one Dirac delta coming from the \# topology also cancel when taking appropriate sums, so we are only left with the $\delta\delta$ structure dictated by \reef{Sexact}.


\small

\bibliography{biblio}
\bibliographystyle{utphys}

\end{document}